\documentclass[aps,prd,reprint,groupedaddress]{revtex4-2}

\usepackage{mathtools}
\usepackage{amsmath}
\usepackage{graphicx}
\usepackage{float}
\usepackage{subcaption}
\usepackage[colorlinks=true, linkcolor=blue, citecolor=blue, urlcolor=blue]{hyperref}

\begin{document}
\title{Strong gravitational lensing in a Kerr black hole within Quantum Einstein Gravity}

\author{Chen-Hao Xie}
\affiliation{Faculty of Science, Kunming University of Science and Technology,\\Kunming 650500, China}

\author{Yu Zhang}
\email{zhangyu\_128@126.com}
\affiliation{Faculty of Science, Kunming University of Science and Technology,\\Kunming 650500, China}

\author{Bo-Li Liu}
\affiliation{Faculty of Science, Kunming University of Science and Technology,\\Kunming 650500, China}

\author{Peng-Fei Duan}
\affiliation{Faculty of Science, Kunming University of Science and Technology,\\Kunming 650500, China}

\author{Yu-Li Lou}
\affiliation{Faculty of Science, Kunming University of Science and Technology,\\Kunming 650500, China}

\date{\today}

\begin{abstract}
The detailed study of the strong gravitational lensing of a Kerr black hole within Quantum Einstein Gravity (QEG) is performed. We calculate the photon sphere, the deflection angle of light, and observables on the equatorial plane under the strong deflection limit in a vacuum. The presence of quantum effects reduces the radius of the photon sphere, the magnification, the position of relativistic images, and the time delays on the same side of the lens. However, it increases the strong deflection angle, the separations, and the time delays on the opposite side of the lens. By modeling M87* and Sgr A* as the Kerr black hole within QEG, we find that the time delays are more significant in M87*, while other observables are more pronounced in Sgr A*. Furthermore, we consider the influence of plasma on the gravitational lensing effect. Plasma causes an additional deflection of light, increasing the magnification, images position and the time delays, but decreasing the separations. More importantly, we calculate the time delays under the strong deflection limit in the presence of plasma, and they increase with higher plasma concentrations. Our research may help to evaluate the observational imprints left by such quantum effects in the propagation of light and the impact of plasma around black holes on gravitational lensing.
\end{abstract}


\maketitle
\section{Introduction}
\label{sec:intro}
From ancient times to the present, people have always pursued the unification of theories. After Einstein completed GR, he began searching for a unified theory of fundamental interactions, i.e., a complete theory of quantum gravity. Since General Relativity (GR) is non-renormalizable \cite{tHooft:1974toh,Birrell:1982ix,Aharony:1998tt}, it is considered the low-energy limit of a certain quantum gravity theory \cite{VanNieuwenhuizen:1981ae}. As a solution, Bonanno and Reuter used the idea of the Wilsonian renormalization group \cite{Wilson:1973jj} to establish a general framework for quantum gravity \cite{Reuter:1996cp}. Additionally, the spherically symmetric black holes \cite{Bonanno:2000ep} and the axisymmetric black holes \cite{Reuter:2010xb,Torres:2017gix} can be derived within the framework.

Black holes are compact astrophysical objects predicted by Einstein's GR. Recent observational data from the Event Horizon Telescope (EHT) collaboration provided convincing evidence for the existence of black holes \cite{EventHorizonTelescope:2019dse,EventHorizonTelescope:2019jan,EventHorizonTelescope:2019pgp,EventHorizonTelescope:2019ggy,EventHorizonTelescope:2022xqj,EventHorizonTelescope:2022wkp,EventHorizonTelescope:2022exc,EventHorizonTelescope:2022urf,Cui:2023uyb}. Black holes are the best laboratories for testing different theories of gravity due to their strong gravitational fields. Specifically, the study of strong gravitational lensing by black holes can reveal the characteristics of strong gravitational fields. Through the observational signatures, we can distinguish between different black holes \cite{Eiroa:2002mk,Gyulchev:2006zg,Bin-Nun:2009hct,Kraniotis:2010gx,Wei:2011nj,Kraniotis:2014paa,Zhao:2016kft,Chakraborty:2016lxo,Zhao:2017cwk,Jusufi:2018jof,Wang:2019cuf,Jusufi:2019caq,Kumar:2020sag,Ghosh:2020spb} and differentiate between various gravitational theories \cite{Bhadra:2003zs,Horvath:2011xr,Sahu:2015dea,Badia:2017art,Allahyari:2019jqz,Li:2020zxi,Afrin:2021imp,Afrin:2021wlj,Kuang:2022ojj,Kuang:2022xjp,Vagnozzi:2022moj,Pantig:2022ely,Soares:2023err}.

According to GR, gravitational lensing occurs when light from a light source passes through a gravitational field, bending as if it were passing through a lens. The degree of light bending mainly depends on the nature of the lens. Gravitational lensing was initially used to test GR in the weak field approximation \cite{Einstein:1936llh}. After that, Darwin \cite{Darwin:1959wai} pioneered research on strong gravitational lensing by compact astrophysical objects, such as black holes. Virbhadra and Ellis obtained the gravitational lens equation in the strong deflection limit and investigated the gravitational lensing by a Schwarzschild black hole \cite{Virbhadra:1999nm}. They showed that when light approaches a black hole, the light source behind the black hole produces a series of images on both sides of the lens, known as relativistic images \cite{Virbhadra:2008ws,Virbhadra:2022iiy,Virbhadra:2022ybp}. Then, Bozza et al. proposed a fully analytical method to define a strong deflection limit by retaining the first two leading order terms in the divergent part. Through this approximation, they developed a method for calculating the deflection angle that is applicable to both spherically symmetric \cite{Bozza:2002zj} and axially symmetric spacetime \cite{Bozza:2002af}, and derived the relevant formulas for the image positions, the separations, the magnification, and the time delays \cite{Bozza:2003cp} between different relativistic images. Nowadays, this method has been widely applied in studies of strong gravitational lensing \cite{Whisker:2004gq,Chen:2009eu,Liu:2010wh,Ding:2010dc,Sotani:2015ewa,Tsukamoto:2016qro,Hsieh:2021scb,Ghosh:2022mka,Tsukamoto:2022tmm,Tsukamoto:2022uoz,AbhishekChowdhuri:2023ekr,Soares:2023uup}. 

The published EHT image of M87* is consistent with the picture that the supermassive black hole in M87 is surrounded by a relativistically hot, magnetized plasma \cite{EventHorizonTelescope:2019dse,EventHorizonTelescope:2019jan,EventHorizonTelescope:2019ggy,EventHorizonTelescope:2019pgp,EventHorizonTelescope:2022xqj,EventHorizonTelescope:2022urf}. Moreover, most astrophysical black holes are surrounded by plasma. Therefore, it is necessary to study the gravitational lensing effect under the influence of plasma. Besides gravity, the plasma around astrophysical objects also significantly affects the motion of photons. Synge presented a general theory of geometrical optics in curved spacetime for an arbitrary medium in his book \cite{Synge:1960ueh}. The comprehensive review of ray optics in media from a general relativistic perspective was developed in the monograph by Perlick \cite{Perlick:2000a}. A series of papers by Bisnovatyi-Kogan and Tsupko provides detailed studies of the physical properties and various scenarios of gravitational lensing in plasma \cite{Bisnovatyi-Kogan:2008qbk,Bisnovatyi-Kogan:2010flt,Tsupko:2013cqa,Tsupko:2014lta,Perlick:2015vta,Bisnovatyi-Kogan:2015dxa,Bisnovatyi-Kogan:2017kii,Tsupko:2019axo,Bisnovatyi-Kogan:2022yzj}. Recently, research on gravitational lensing by different black holes filled with plasma has been flourishing \cite{Er:2013efa,Liu:2016eju,Turimov:2018ttf,Hensh:2019ipu,Jin:2020emq,Atamurotov:2022slw,Atamurotov:2022xvs,Er:2022lad,Li:2023esz}.

Many physical phenomena related to the Kerr black hole within QEG have been explored, such as the shadow of the black hole and the weak deflection limit \cite{Kumar:2019ohr}. In the research of the black hole shadow, the authors investigated the shadows cast by rotating black holes using the Hamilton-Jacobi equation and the Carter separable method. It turns out that the size of black hole shadows monotonically decreases and shadows get more distorted with increasing values of additional parameters (except $M$ and $a$) when compared with the Kerr black hole shadows. In this paper, we will explore the influence of additional parameters and plasma on gravitational lensing under the strong deflection limit for the Kerr black hole within QEG using the method of Bozza \cite{Bozza:2002zj,Bozza:2002af}.

The paper is structured as follows. Section~\ref{Spacetime} briefly reviews the Kerr black hole within QEG. In section \ref{Strong gravitational deflection}, we study the gravitational lensing effect in the strong deflection limit and calculate the strong deflection angle. Section~\ref{Observables} models M87* and Sgr A* as Kerr black holes within QEG to calculate the strong lensing observables and evaluates their deviation from the Kerr situation. The strong gravitational lensing effect in the plasma is studied in section~\ref{Strong gravitational deflection in plasma}. Finally, we conclude and discuss our results in section~\ref{Conclusion}.

\section{Spacetime}
\label{Spacetime} 
Reuter proposed a general framework for dealing with quantum gravity along the lines of Wilson's renormalization group \cite{Reuter:1996cp}. In a nutshell, he introduced a scale-dependent effective average action \cite{Wetterich:1992yh} and derived an exact renormalization group equation \cite{Reuter:1996cp}. Then, he obtained the running Newton's constant \cite{Reuter:1996cp} from the exact evolution equation for the effective average action. In ref.\cite{Bonanno:2000ep}, Bonanno and Reuter constructed a renormalization group-improved quantum gravity Schwarzschild black hole, based on the running Newton's constant and the quantum improved black hole spacetime coming from QEG. Subsequently, the authors extended it to rotating black holes, the metric of which is \cite{Reuter:2010xb,Torres:2017gix}
\begin{align}\label{metric}
ds^2=&-\left(1-\frac{2 M G(k) r}{\rho}\right)dt^2+\frac{\rho^2}{\Delta}dr^2+\rho^2d\theta^2 \notag \\ 
&+\frac{\Sigma\sin^2\theta}{\rho^2}d\varphi^2-\frac{4 M G(k) r a \sin^2 \theta}{\rho^2}dtd\varphi ,
\end{align}
where
\begin{align}\label{pa}
     &\rho^2=r^2+a^2\cos^2\theta, \\ &\Delta=r^2-2M G(k) r+a^2, \\ &\Sigma=(r^2+a^2)^2-\Delta a^2\sin^2\theta.
\end{align}

In the QEG approach, through the use of the functional renormalization group equation, the authors found that the $k$ dependence of the running Newton's constant $G(k)$ can be given approximately by \cite{Reuter:1996cp,Bonanno:2000ep,Reuter:2010xb,Torres:2017gix}
\begin{equation}
    G(k) = \frac{G_0}{1 + \omega G_0 k^2},
\end{equation}
where $G_0$ is the classical gravitational constant and $\omega$ is a constant. After that, in order to convert the scale dependence of the Newton's constant into position dependence, the energy scale can be written as \cite{Bonanno:2000ep,Reuter:2010xb,Torres:2017gix}
\begin{equation}
    k(P) = \frac{\xi}{d(P)},
\end{equation}
where $\xi$ is a constant to be determined later and $d(P)$ is the distance scale which provides the relevant cutoff for the Newton's constant when a test particle is located at a point $P$ in black hole spacetime.

If we use Boyer-Lindquist coordinates ($t$, $r$, $\theta$, $\phi$) and consider axially symmetric spacetime, then this implies that $d(P)$ only depends on the $r$-coordinate and $\theta$-coordinate of $P$, i.e., $d = d(r,\theta)$. A specific expression for the angular dependence was not found \cite{Torres:2017gix}, and this paper only considers the equatorial plane, so $G(k)$ can be written as \cite{Reuter:2010xb,Torres:2017gix}
\begin{align}
    G(r)=\frac{G_0d^2(r)}{d^2(r)+{\tilde{\omega}}G_0},
\end{align}
where $\tilde{\omega} = \omega \xi^2 $.

For the choice of $d(r)$, $d(r) = r$ is ``naive'' \cite{Bonanno:2000ep}, but it only softens the singularity. In contrast, choosing $d(r)=\left(\frac{r^3}{r+\gamma G_0 M}\right)^{\frac{1}{2}}$ \cite{Bonanno:2000ep,Bonanno:2006eu,Torres:2017gix} can completely eliminate the central singularity. Therefore, the final form of $G(r)$ we have chosen is \cite{Bonanno:2000ep,Bonanno:2006eu,Torres:2017gix}: 
\begin{equation}\label{Gr}
    G(r)=\frac{G_0 r^3}{r^3+{\tilde{\omega}} G_0 (r+\gamma G_0M)},
\end{equation}
where ${\tilde{\omega}}$ and $\gamma$ are come from the non-perturbative renormalization group theory and the suitable cutoff identification, respectively. Among them, the existence of ${\tilde{\omega}}$ symbolizes the quantum corrections in the classical black hole, and its value determines the strength and nature of these corrections. The preferred theoretical value of $\gamma$ is $\gamma=9/2$ \cite{Bonanno:2000ep,Bonanno:2006eu,Torres:2017gix}, from which the value of ${\tilde{\omega}}$ can be inferred to be $\frac{167\hbar}{30\pi}$ \cite{Torres:2017gix}. In fact, the properties of the solutions do not depend on their exact values, as long as they are strictly positive \cite{Bonanno:2000ep,Torres:2017gix}. By setting $\gamma = 0$, we can return to the ``naive'' cutoff. Furthermore, when ${\tilde{\omega}}=0$, the switch for the quantum correction will be turned off, and $G(r)$ reverts to $G_0$, i.e., the metric returns to the Kerr metric.

\begin{figure}[t]
\centering
\includegraphics[width=8.6cm]{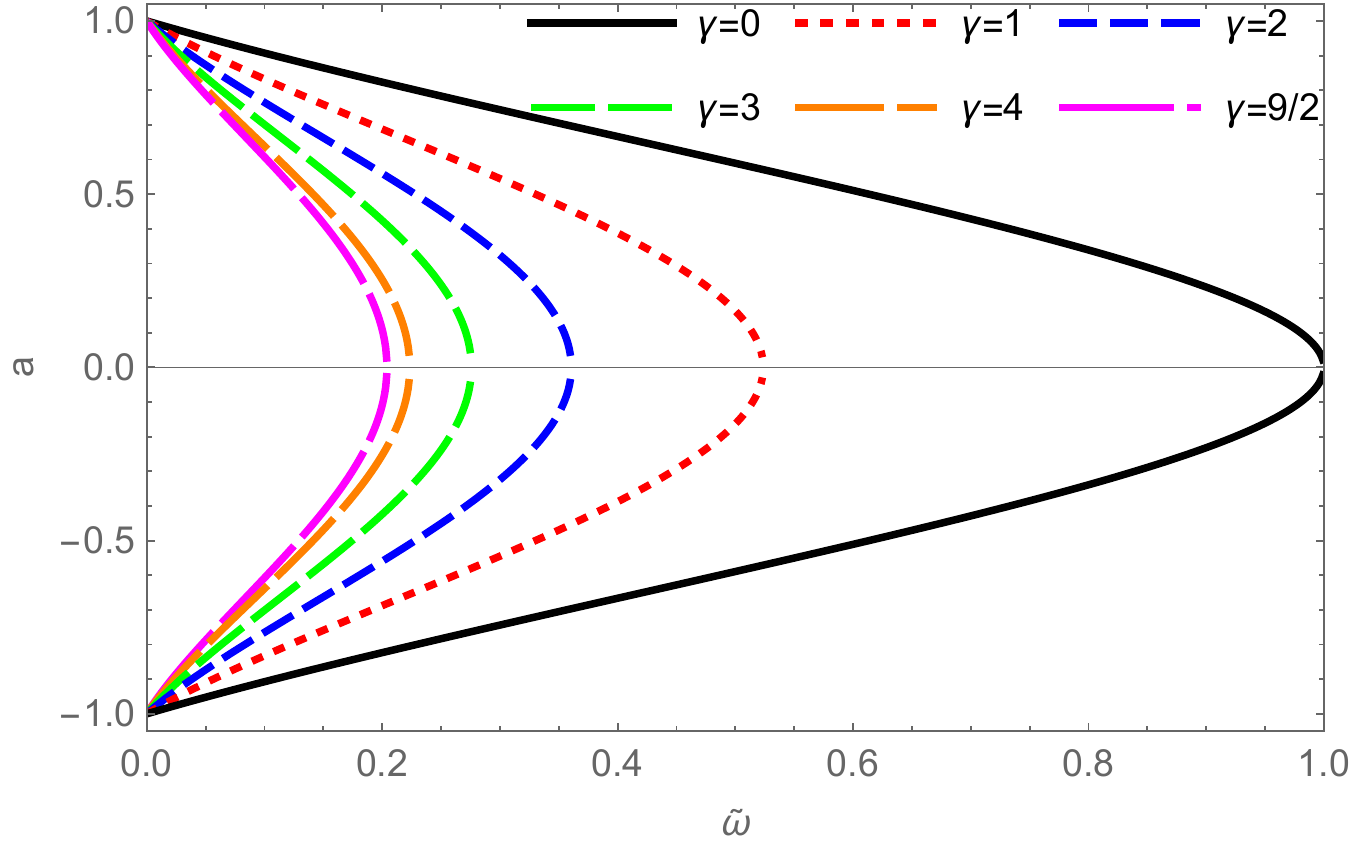}
\caption{The existence region of black holes for different $a$, $\tilde{\omega}$ and $\gamma$. The left side of the line represents the presence of a black hole, while the right side represents the absence of a black hole.}
\label{Region}
\end{figure}

In conclusion, eqs.~(\ref{metric}),~(\ref{pa}), and~(\ref{Gr}) express the spacetime metric discussed in this article. In this paper, we consider $\gamma$ as a free parameter, and fig.~\ref{Region} shows the regions where black holes exist for different ${\tilde{\omega}}$, $\gamma$, and $a$ (All instances of $a < 0$ in this paper represent the photon's direction of motion being opposite to the direction of the black hole's rotation). We can see that both ${\tilde{\omega}}$ and $\gamma$ will cause the black hole to become an extreme black hole before $a$ approaches $M$.

\section{Strong gravitational deflection}
\label{Strong gravitational deflection}
In this section, we discuss the strong gravitational deflection of photon rays in Kerr spacetime within QEG. The equatorial motion can yield very obvious indications and help to understand the general case, so we consider the deflection of photons in the equatorial plane ($\theta = \pi/2$) and in the units of $M=1$. The reduced metric has the form:
\begin{equation}\label{simmetric}
    d s^{2}=-A(r) d t^{2} + B(r) d r^{2} + C(r)d \phi ^{2} - D(r) d t d \phi.
\end{equation}

In the presence of a gravitational field, the trajectories of photons may be obtained from the variational principle \cite{Synge:1960ueh}:
\begin{equation}\label{variational principle}
    \delta\left(\int p_idx^i\right)=0,
\end{equation}
and Hamiltonian of the photons around the Kerr black hole within QEG has the following \cite{Kulsrud:1991jt}:
\begin{equation}
    H(x^i,p_i)=\frac{1}{2}g^{ik}p_ip_k=0.
\end{equation}

Through the variational principle~(\ref{variational principle}) and $H(x^i,p_i)=0$, we obtain the Hamiltonian differential equations \cite{Synge:1960ueh}:
\begin{equation}\label{Hdeq}
    \frac{dx^i}{d\lambda}=\frac{\partial H}{\partial p_i} , \frac{dp_i}{d\lambda}=-\frac{\partial H}{\partial x^i},
\end{equation}
with the affine parameter $\lambda$ changing along the light trajectory. Then we get two constants of motion, which are the energy and angular momentum of the particle:
\begin{equation}\label{Const}
    E=-p_t,\quad L=p_\phi.
\end{equation}

We can set $E=1$ by choosing an appropriate affine parameter and replace $L/E$ with the impact parameter $u$. From eqs.~(\ref{Hdeq}) and~(\ref{Const}), the expressions for the motion equations of photons can be obtained as follows:
\begin{align}
        \frac{dt}{d\lambda}& = \frac{4 C(r)- D(r)u}{D(r)^{2}+4 A(r)C(r)}, \label{dt} \\
        \frac{d\phi}{d\lambda}& = \frac{2 D(r)+4  A(r)u}{D(r)^{2}+4 A(r)C(r)}, \label{dphi} \\
        \left(\frac{dr}{d\lambda}\right)^{2}& = \frac{4 C(r)- 4 D(r) u - 4 A(r)u^{2}}{B(r)[D(r)^{2}+4 A(r)C(r)]}. \label{dr}
\end{align}

The effective potential for radial motion of photons can be defined as
\begin{equation}\label{effp}
    V_{\mathrm{eff}}:=-\dot{r}^{2}=-\frac{4(C-Du-Au^{2})}{B(4AC+D^{2})}.
\end{equation}

The effective potential characterizes different types of photon orbits. In the process of photons being deflected by gravity, photons approach the black hole from infinity, reach the closest distance, and then shoot towards infinity again. The photon's effective potential disappears ($V_{\mathrm{eff}}:=-\dot{r}^{2} = 0$) at the closest approach distance $r_0$, so we can obtain the relationship between the closest approach distance $r_0$ and the impact parameter through this condition as follows:
\begin{equation}\label{impactp}
    L=u(r_0)=\frac{-D_0+\sqrt{D_0^2+4A_0C_0}}{2A_0},
\end{equation}
where the subscript $0$ implies that the function is evaluated at $r = r_0$. We choose the positive sign before the square root, and the positive and negative values of $a$ represent the prograde and retrograde motion of the photons. As the light rays approach the photon sphere, the deflection angle will tend to infinity at $r_0 = r_m$, which is determined by the following equation:
\begin{equation}\label{condition}
    V_{\text{eff}}=\left.\frac{dV_{\text{eff}}}{dr}\right|_{r_0=r_m}=0.
\end{equation}

\begin{figure*}[t]
\centering
\begin{subfigure}[b]{8.6cm}
        \centering
        \includegraphics[width=8.6cm]{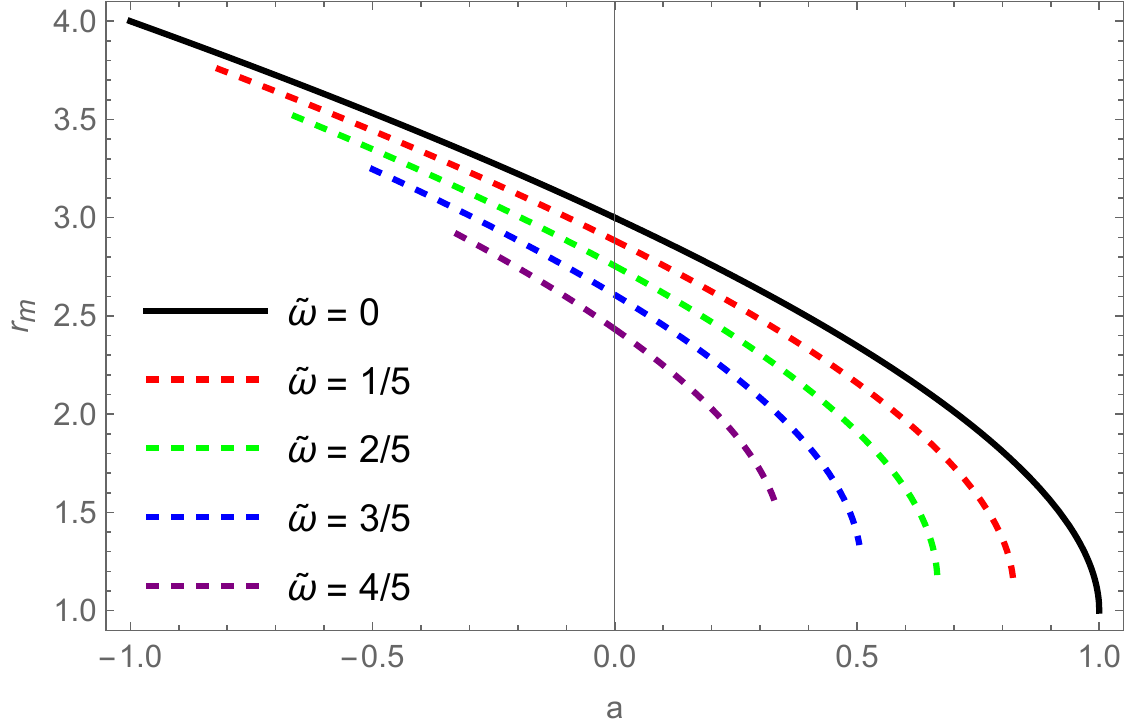}
    \end{subfigure}
    \hspace{0.01\linewidth}
    \begin{subfigure}[b]{8.6cm}
        \centering
        \includegraphics[width=8.6cm]{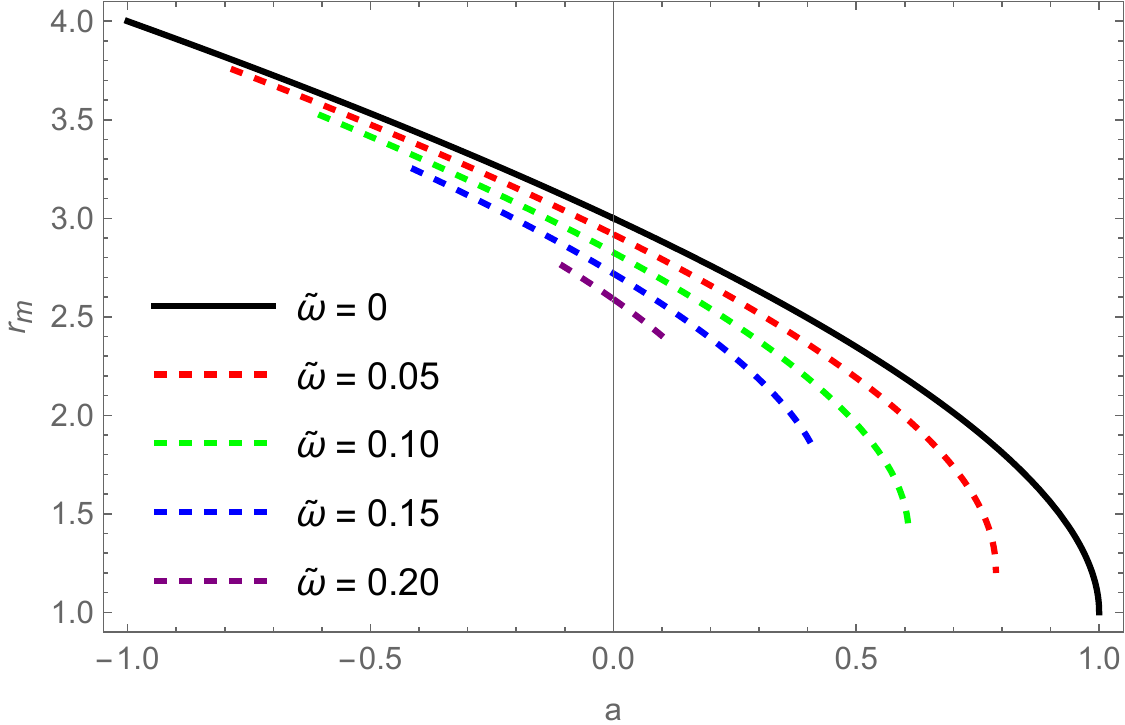}
    \end{subfigure}
    \caption{The behavior of the photon sphere radius $r_m$ as a function of  $a$ for the Kerr black hole within QEG with different values of  $\tilde{\omega}$. The left panel present $\gamma = 0$ and the right panel present $\gamma = 9/2$.}
    \label{Rm}
\end{figure*}

Therefore, from eqs.~(\ref{effp}),~(\ref{impactp}), and~(\ref{condition}), we can obtain the photon sphere equation as
\begin{equation}
    AC'-A'C+u(r_0)(A'D-AD')=0,
\end{equation}
and the largest root of it is the photon sphere radius $r_m$. In fig.~\ref{Rm}, the photon sphere radius decreases as the spin $a$ increases for different $\tilde{\omega}$. Moreover, the left and right panels respectively show the cases of $\gamma = 0$ and $\gamma = 9/2$. The results indicate that the radius $r_m$ decreases with the increase of $\tilde{\omega}$, and for the same $\tilde{\omega}$, the presence of $\gamma$ makes $r_m$ smaller. In the entire following content, we only consider the case where $\gamma = \frac{9}{2}$.

Based on the radius of the photon sphere, we can begin calculating the strong deflection angle. The deflection angle of the light rays corresponding to $r_0$ can be obtained by combining eqs.~(\ref{dphi}) and~(\ref{dr}):
\begin{align}
    \alpha(r_0)&=\phi(r_0)- \pi, \label{angle1}\\
    \phi(r_0) &= 2\int_{r_{0}}^{\infty}\frac{d\phi}{dr}dr. \label{phi}
\end{align}

Along the lines of Bozza's method \cite{Bozza:2002zj,Bozza:2002af}, we can obtain the approximate analytical form of the light deflection angle close to $r_m$. First, we define a new variable $z$ as
\begin{equation}
   z = \frac{A - A_0}{1 - A_0},
\end{equation}
and replace the variable $r$ with $z$, so eq.~(\ref{phi}) can be rewritten as
\begin{align}
    \phi(r_{0})& =\int_{0}^{1}R(z,r_{0})f(z,r_{0})dz, \label{integralPhi} \\
    R(z,r_{0})& =2\frac{(1-A_{0})}{A'}\frac{\sqrt{B |A_0|}(2Au+D)}{\sqrt{A_{0}C}\sqrt{4AC+D^{2}}}, \\
    f(z,r_{0})& =\frac{1}{\sqrt{\frac{\mathrm{sgn}(A_0)}{C}\bigl[ A_0 C-AC_0+u(AD_0-A_0D)\bigr]}}. \label{f}
\end{align}

The function $R(z,r_0)$ is regular everywhere, while $f(z,r_0)$ diverges as $z \to 0$. Thus, the integral can be separated into the divergent part and the regular part
\begin{equation}
    \begin{aligned}
        &\phi_{D}(r_{0})=\int_{0}^{1}R(0,r_{\mathrm{m}})f_{0}(z,r_{0})dz,\\
        &\phi_{\mathrm{R}}(r_{0})=\int_{0}^{1}[R(z,r_{0})f(z,r_{0})-R(0,r_{0})f_{0}(z,r_{0})]dz.
    \end{aligned}
\end{equation}

In order to find out the order of divergence of the integrand, the argument of the square root in $f(z,r_0)$ is expanded using the Taylor series to the second order in $z$:
\begin{equation}
    f(z,r_0)\sim f_0(z,r_0)=\frac{1}{\sqrt{{\alpha}z+{\beta}z^2}},
\end{equation}
where
\begin{equation}
   \begin{aligned}
        \alpha &=\mathrm{sgn}(A_0)\frac{(1-A_0)}{A_0^{\prime}C_0}(A_0C_0^{\prime}-A_0^{\prime}C_0+u (A_0^{\prime}D_0-A_0D_0^{\prime})), \\
        \beta&=\operatorname{sgn}(A_0)\frac{(1-A_0)^2}{2C_0^2A_0^{\prime3}}\bigg\{2C_0C_0'A_0'^2+(C_0C_0''-2C_0'^2)A_0A_0'  \\
        & -C_0C_0'A_0A_0'' +u[A_0C_0(A_0''D_0'-A_0'D_0'')\\
        & +2A_0'C_0'(A_0D_0'-A_0'D_0)]\bigg\}.
    \end{aligned}
\end{equation}

For the case where $r_0$ is close to $r_m$, i.e., in the strong deflection limit, we can get an analytical expansion of the deflection angle close to the divergence in the form \cite{Bozza:2002zj,Bozza:2002af,Tsukamoto:2016jzh}
\begin{align}
    \alpha_{Sd}(\theta)=&-\bar{a}\log\left(\frac{\theta D_{OL}}{u_m}-1\right)+\bar{b} \notag \\
    &+\mathcal{O}\boldsymbol{(}(u-u_m)\log(u-u_m)\boldsymbol{)},
\end{align}
where $\bar{a}$ and $\bar{b}$ are strong deflection limit coefficients. According to the above relationship, we can give them as follows:

\begin{align}
    &\bar{a}=\frac{R(0,r_{m})}{2\sqrt{{\beta}_{m}}},\\&\bar{b}=-\pi+\bar{b}_{D}+\bar{b}_{R}+\bar{a}\log\frac{cr_{m}^{2}}{u_{m}},
\end{align}

where
\begin{equation}
    \begin{aligned}
        &\bar{b}_{D}=2\bar{a}\log\frac{2(1-A_m)}{A_m^{\prime}r_m},\\&\bar{b}_{R}=\int_0^1\left[R(z,r_m)f(z,r_m)-R(0,r_m)f_0(z,r_m)\right]dz,
    \end{aligned}
\end{equation}
and $c$ is defined by
\begin{equation}
    u(r_0) -u(r_m)=c(r_0-r_m)^2.
\end{equation}

\begin{figure*}[t]
\centering
\begin{subfigure}[b]{8.6cm}
        \centering
        \includegraphics[width=8.4cm]{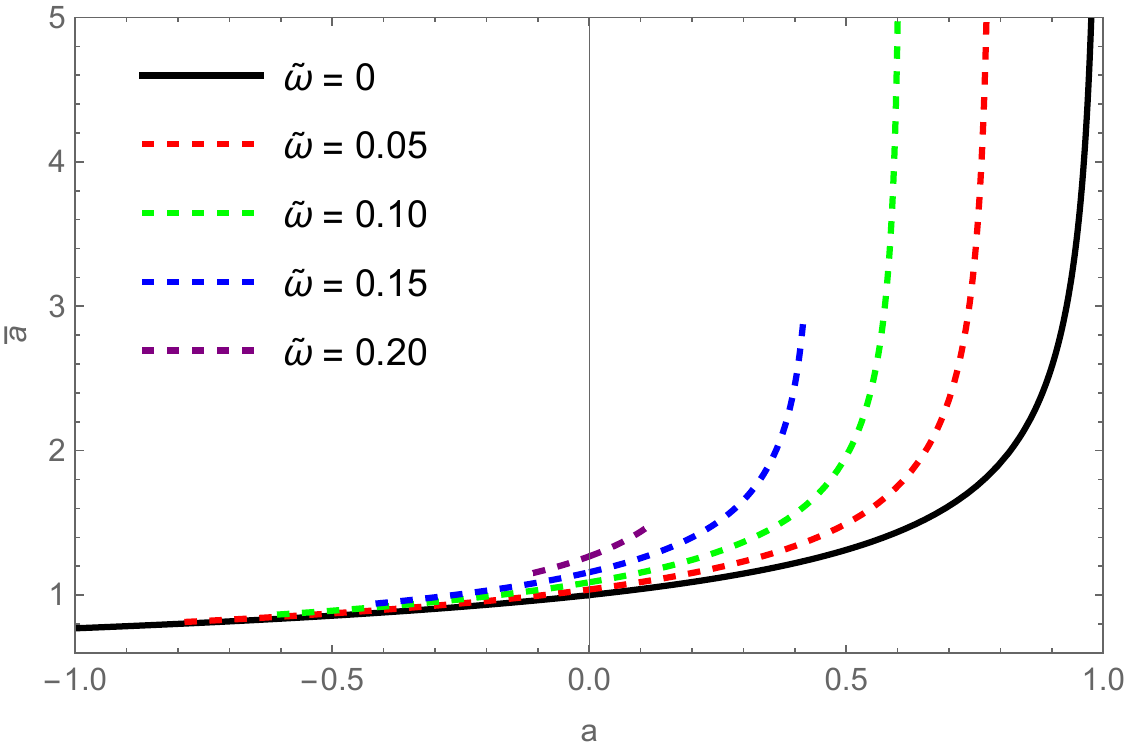}
    \end{subfigure}
    \hspace{0.01\linewidth}
    \begin{subfigure}[b]{8.6cm}
        \centering
        \includegraphics[width=8.7cm]{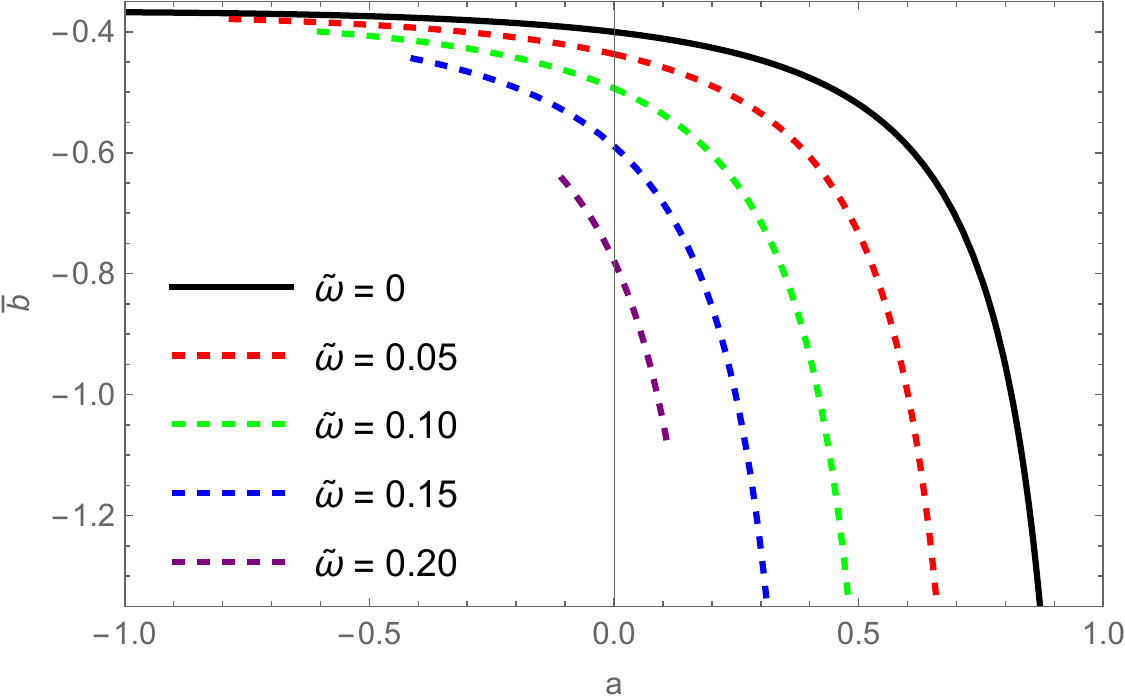}
    \end{subfigure}
    \caption{The behaviors of the strong deflection limit coefficients $\bar{a}$ and $\bar{b}$ as functions of $a$ for the Kerr black hole within QEG with different values of $\tilde{\omega}$.}
    \label{ab}
\end{figure*}

\begin{figure}[t]
\centering
\includegraphics[width=8.6cm]{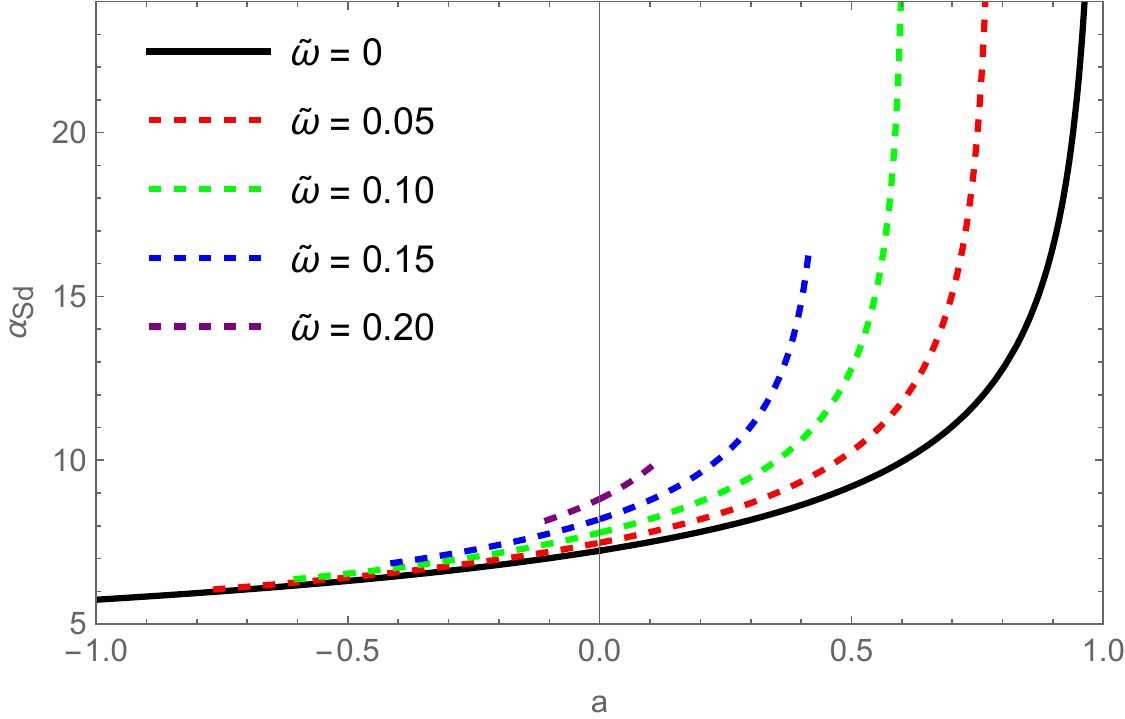}
\caption{The deflection angle $\alpha_{Sd}$ as the a function of $a$ for the Kerr black hole within QEG with different $\tilde{\omega}$ when $u = u_m + 0.0025$.}
\label{angle}
\end{figure}

Based on the results obtained above, we can study the deflection angle of the Kerr metric within Quantum Einstein Gravity under the strong deflection limit and compare it with the Kerr case. Fig.~\ref{ab} depicts the strong deflection limit coefficients $\bar{a}$ and $\bar{b}$ as functions of the spin $a$ for different $\tilde{\omega}$. We can see that $\bar{a}$ increases with the increase in $a$ or $\tilde{\omega}$, and $\bar{b}$ decreases with the increase in $a$ or $\tilde{\omega}$. Furthermore, we plot the behavior of the strong deflection angle $\alpha_{Sd}$ for the Kerr black hole within QEG in fig.~\ref{angle}. However, $\tilde{\omega}$ increases the deflection angle $\alpha_{Sd}$ of light when $u = u_m + 0.0025$ (near the photon sphere). This means that the presence of $\tilde{\omega}$, namely the presence of the quantum effects reduces the radius of the photon sphere in the strong deflection limit, but it increases the strong deflection angle.

\section{Observables}
\label{Observables}
The optimal scenario for observing relativistic images occurs when the source is nearly aligned with the lens, which we will investigate in this section. We assume that both the source and the observer are far from the lens and perfectly aligned in an asymptotically flat spacetime. Therefore, the lens equation in the strong deflection limit reads \cite{Bozza:2001xd,Bozza:2008ev}
\begin{equation}\label{lenseq}
    \beta_{LS}=\theta-\frac{D_{LS}}{D_{OS}}\Delta\alpha_{n},
\end{equation}
where $\Delta\alpha_n=\alpha_{Sd}(\theta)-2n\pi$ is the offset of the deflection angle, $D_{LS}$ measures the distance from the lens to the source, $D_{OS} = D_{OL}+D_{LS}$, $D_{OL}$ is the distance from observer to the lens, $\beta_{LS}$ represents the angular separation between the lens and the source, and $\theta$ is the angular separation between the lens and the image. Starting from the lens equation and the deflection angle equation, we can obtain the positions of relativistic images, the separations, the magnification, and the time delays between different images.

Firstly, expanding the deflection angle $\alpha(\theta)$ at $\theta = \theta^{0}_{n}$ and setting $\Delta\theta_n=\theta-\theta_n^0$, we get \cite{Bozza:2002zj}
\begin{equation}
    \Delta\alpha_n=-\frac{\overline{a}D_{OL}}{u_m e_n}\Delta\theta_n.
\end{equation}

Here, $\theta_{n}^{0}$ is the image position corresponding to $\alpha(\theta_{n}^{0})=2n\pi$. This means
\begin{equation}\label{position1}
    \theta_n^0=\frac{u_m}{D_{OL}}\left(1+e_n\right),
\end{equation}
where $e_n=\exp\left(\frac{\bar{b}-2n\pi}{\bar{a}}\right)$.

Combining eq.~(\ref{lenseq}) and eq.~(\ref{position1}), we can approximately obtain the position of the $n$th relativistic image:
\begin{equation}\label{position2}
    \theta_n\simeq\theta_n^0+\frac{u_m e_n(\beta_{LS}-\theta_n^0)D_{OS}}{\bar{a}D_{LS}D_{OL}}.
\end{equation}

The gravitational lensing effects will magnify the brightness of the source image, and the image magnification is defined as the ratio between the solid angles of the relativistic image and the source. Therefore, the magnification of the $n$th relativistic image can be given by \cite{Bozza:2002zj}
\begin{equation}\label{mag}
    \mu_n=\left.\left(\frac{\beta_{LS}}{\theta}\frac{d\beta_{LS}}{d\theta}\right)^{-1}\right|_{\theta_n^0}=\frac{u_m^2e_n(1+e_n)D_{OS}}{\bar{a}\beta_{LS}D_{LS}D_{OL}^2},
\end{equation}
where $\mu_n$ decreases exponentially with $n$, making the images are fainter as $n$ increases. In addition, the magnification is proportional to $1/D_{OL}^2$, so the image is relatively bright when $\beta_{LS} \to 0$ (perfect alignment).

Eqs.~(\ref{position2}) and~(\ref{mag}) relate the images positions and the magnification to the strong deflection limit coefficients. Next, we will consider the inverse problem, i.e., obtaining the strong deflection limit coefficients that carry black hole information from the position and the magnification. To effectively relate the observables to the strong deflection limit coefficients, we treat the image $\theta_1$ as a single image. All the remaining images are packed together as $\theta_{\infty}$, which is obtained by eq.~(\ref{position2}) in the limit $n \to \infty$. Then, we obtain the positions of the remaining inner packed images $\theta_{\infty}$, the separations between the first image and the others $s$, and the magnification $r_{mag}$ of the first image to the other images \cite{Bozza:2002zj}:
\begin{equation}
    \begin{aligned}
        \theta_{\infty}& =\frac{u_{m}}{D_{OL}}, \\
        &s=\theta_{1}-\theta_{\infty}=\theta_{\infty}e^{\frac{\bar{b}-2\pi}{\bar{a}}}, \\
        &r =\frac{\mu_{1}}{\sum_{n=2}^{\infty}\mu_{n}}=e^{\frac{2\pi}{\bar{a}}},\quad r_{\mathrm{mag}}\simeq2.5\log_{10}(r)=\frac{5\pi}{\bar{a}\ln10}.
    \end{aligned}
\end{equation}

So far, we link the observables with the strong deflection limit coefficients. This allows us to predict astronomical observations based on the properties of the black hole or infer the nature of the black hole based on astronomical observations.

Furthermore, if the time signals of the different images can be distinguished, it allows us to consider another significant observable in the strong deflection lensing: the time delays. In the multiple relativistic images formed by gravitational lensing, the travel time of light paths corresponding to different relativistic images generally varies. Therefore, there are time delays between different relativistic images, which is determined by the nature of the black hole, the position of the observer, and the position of the light source. If we consider two photons traveling along different trajectories, this results in the time delays between two relativistic images given by \cite{Bozza:2003cp}
\begin{equation}\label{timedelay}
    \begin{aligned}
        T_{1}-T_{2}=&\widetilde{T}(r_{0,1})-\widetilde{T}(r_{0,2})+2\int_{r_{0,1}}^{r_{0,2}}\frac{\widetilde{P}_{1}(r,r_{0,1})}{\sqrt{A_{0,1}}}dr\\
        & +2\int_{r_{0,2}}^\infty\left[\frac{\widetilde{P}_1(r,r_{0,1})}{\sqrt{A_{0,1}}}-\frac{\widetilde{P}_1(r,r_{0,2})}{\sqrt{A_{0,2}}}\right]dr
    \end{aligned}
\end{equation}
with
\begin{align}
    \widetilde{T}(r_0)& =\int_0^1\widetilde{R}(z,r_0) f(z,r_0) dz, \label{integralT} \\
    \widetilde{R}(z,r_0)& =2\frac{1-A_{0}}{A^{\prime}(r)}\widetilde{P}_{1}(r,r_{0})\left(1-\frac{1}{\sqrt{A_{0}} f(z,r_{0})}\right),\\
    \widetilde{P}_{1}(r,r_{0})& =\frac{\sqrt{BA_{0}}(2C-uD)}{\sqrt{C}\sqrt{4AC+D^{2}}},
\end{align}
and $f(r,r_0)$ is given by eq.~(\ref{f}). Here, $r_{0,1}$ and $r_{0,2}$ represent the closest approach distance distance between the two photons and the black hole, respectively. The method for handling the integral~(\ref{integralT}) is similar to that for integral~(\ref{integralPhi}), except that $R(z,r_0)$ needs to be replaced with $\tilde{R}(z,r_0)$. The result is as follows \cite{Bozza:2003cp}:
\begin{equation}
    \widetilde{T}(u)=-\tilde{a}\log\left(\frac{u}{u_m}-1\right)+\tilde{b}+O\left(u-u_m\right)
\end{equation}
with
\begin{align}
    &\tilde{a}=\frac{\widetilde{R}(0,r_m)}{2\sqrt{\bar{\beta}_m}}, \\
    &\tilde{b} = -\pi+\tilde{b}_{D}\left(r_{m}\right)+\tilde{b}_{R}\left(r_{m}\right)+\tilde{a}\log\left(\frac{cr_{m}^{2}}{u_{m}}\right) , \\
    &\tilde{b}_{D}(r_{m}) =2\tilde{a}\log\frac{2(1-A_{m})}{A_{m}^{\prime}r_{m}}, \\
    &\tilde{b}_{R}(r_{m}) =\int_{0}^{1}[\widetilde{R}(z,r_{m})f(z,r_{m})-\widetilde{R}(0,r_{m})f_{0}(z,r_{m})]dz,
\end{align}
where $u_m$ is the critical impact parameter.

Now, we can give the time delays between the $n$th and $m$th images on the same side of the lens as follows \cite{Bozza:2003cp}:
\begin{align}\label{TD1}
     \Delta T_{n,m}^s & \approx 2\pi(n-m)\frac{\tilde a}{\bar a}  \notag \\ 
    & +2\sqrt{\frac{A_m u_m}{B_m}}\left[e^{(\bar b-2m\pi\pm\beta_{LS})/2\bar a} - e^{(\bar b-2n\pi\pm\beta_{LS})/2\bar a}\right].
\end{align}

For the same side situation, the other terms contribute a small proportion, so we only consider the dominant term, and the expression reads as \cite{Bozza:2003cp}
\begin{equation}
    \Delta T_{n,m}^{s}=2\pi(n-m)\frac{\tilde{a}}{\bar{a}}.
\end{equation}

Due to the different travel times of prograde and retrograde photons in rotating spacetime, the time delays on the opposite sides of the lens differ. The dominant term of the time delays for the opposite sides situation is given by the following \cite{Bozza:2003cp}:
\begin{align}
    \Delta\widetilde{T}_{n,m}^o & \approx  \frac{\tilde{a}(a)}{\bar{a}(a)}[2\pi n+\beta_{LS}-\bar{b}(a)]+\tilde{b}(a) \notag \\ 
    &-\frac{\tilde{a}(-a)}{\bar{a}(-a)}[2\pi m-\beta_{LS}-\bar{b}(-a)]-\tilde{b}(-a).
\end{align}

\begin{figure*}[t]
\centering
\begin{subfigure}[b]{8.6cm}
        \centering
        \includegraphics[width=8.7cm]{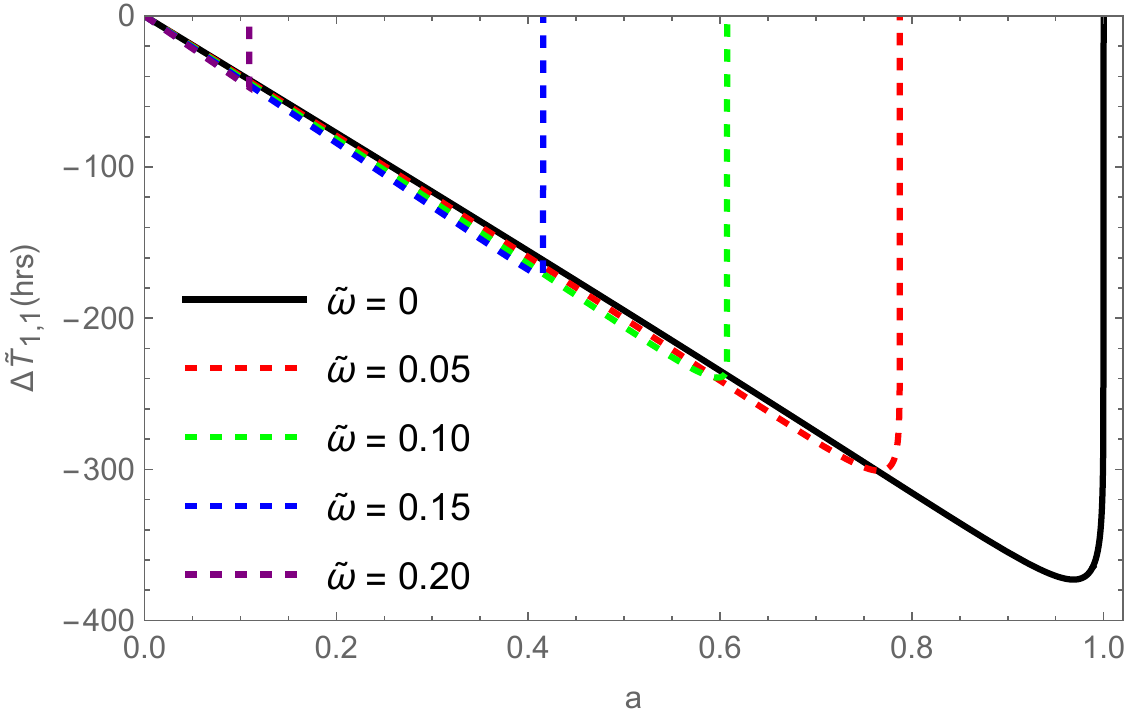}
    \end{subfigure}
    \hspace{0.01\linewidth}
    \begin{subfigure}[b]{8.6cm}
        \centering
        \includegraphics[width=8.5cm]{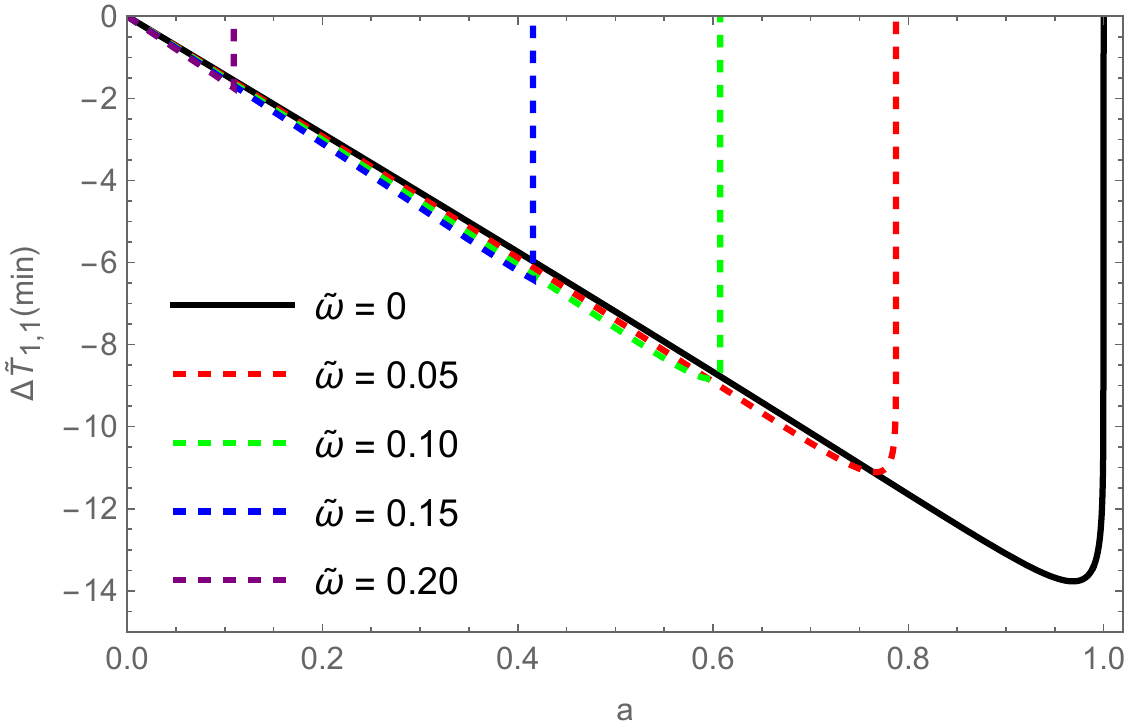}
    \end{subfigure}
    \caption{The behavior of the time delays of first prograde and first retrograde images on the opposide side $\Delta \widetilde{T}_{1,1}$ as a function of $a$ by model M87* (left panel) and Sgr A* (right panel) as the Kerr black hole within QEG.}
    \label{ioTD2}
\end{figure*}

However, it is worth noting that for the time delays on the opposite side, the second integral in eq.~(\ref{timedelay}) is of the same order as the leading term, so we consider all terms in eq.~(\ref{timedelay}), as shown in fig.~\ref{ioTD2}.

\begin{figure}[t]
\centering
\includegraphics[width=8.6cm]{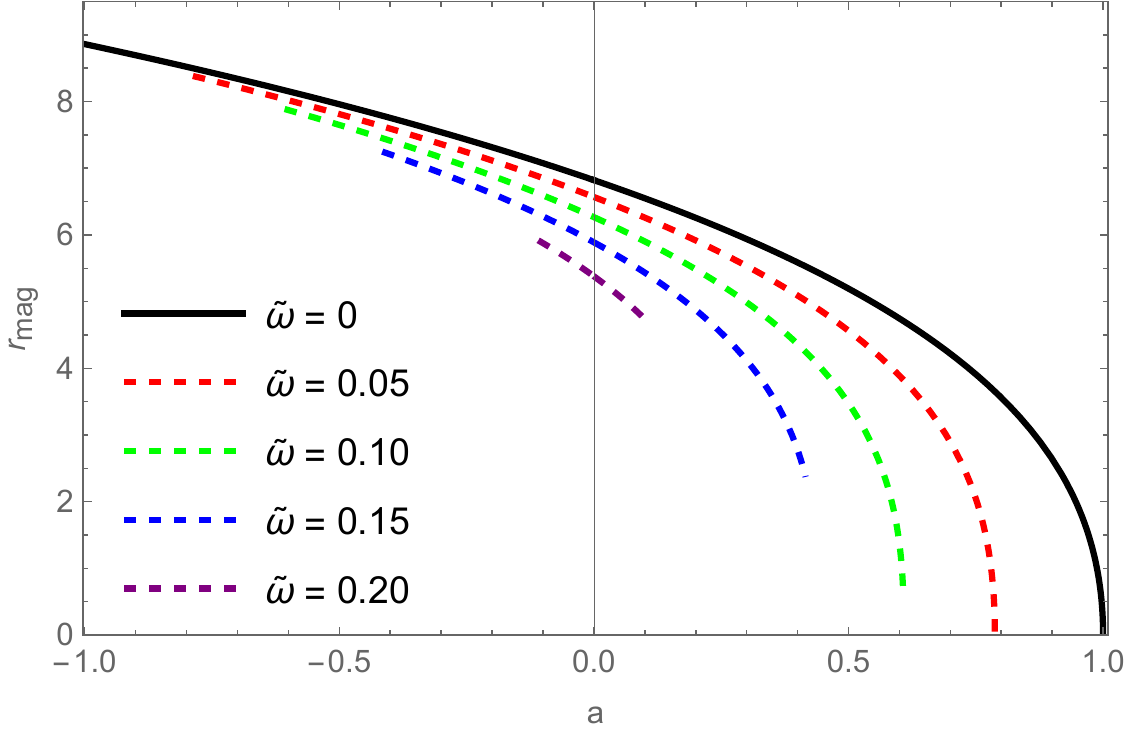}
\caption{The behavior of the magnification $r_{mag}$ as a function of $a$ in the strong deflection limit.}
\label{rmag}
\end{figure}

\begin{figure*}[t]
\centering
\begin{subfigure}[b]{8.6cm}
        \centering
        \includegraphics[width=8.6cm]{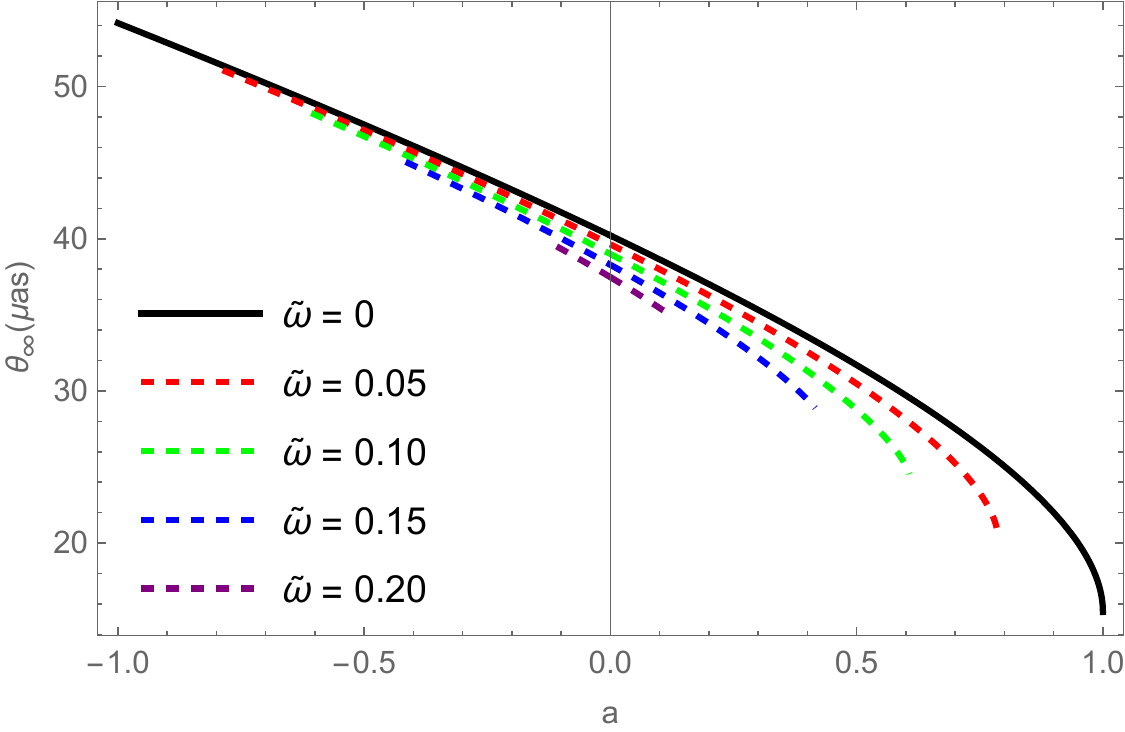}
    \end{subfigure}
    \hspace{0.01\linewidth}
    \begin{subfigure}[b]{8.6cm}
        \centering
        \includegraphics[width=8.6cm]{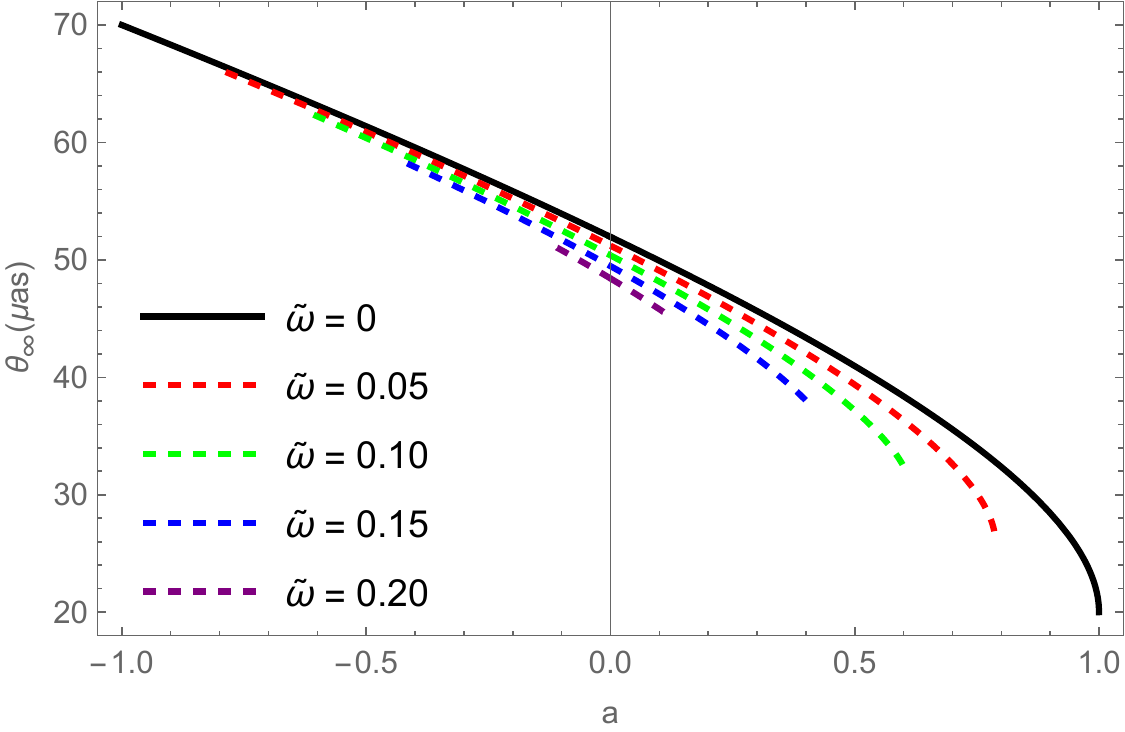}
    \end{subfigure}
    \caption{The behavior of the relativistic image position $\theta_{\infty}$ as a function of $a$ by model M87* (left panel) and Sgr A* (right panel) as the Kerr black hole within QEG.}
    \label{theta}
\end{figure*}

\begin{figure*}[t]
\centering
\begin{subfigure}[b]{8.6cm}
        \centering
        \includegraphics[width=8.6cm]{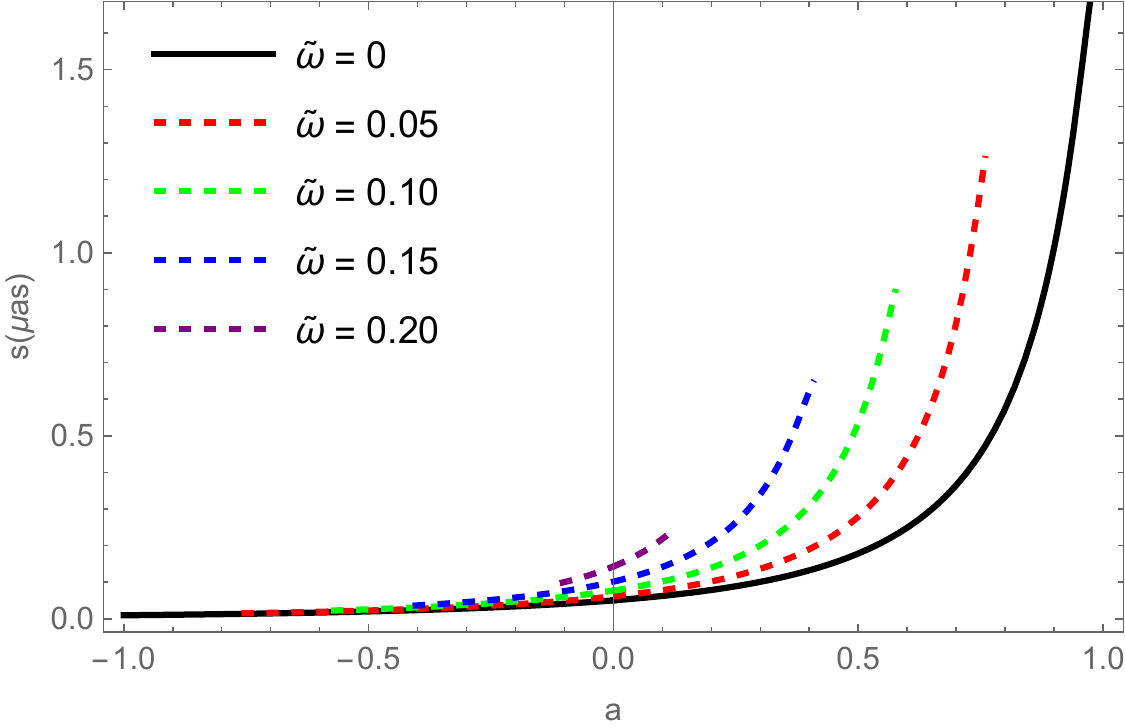}
    \end{subfigure}
    \hspace{0.01\linewidth}
    \begin{subfigure}[b]{8.6cm}
        \centering
        \includegraphics[width=8.6cm]{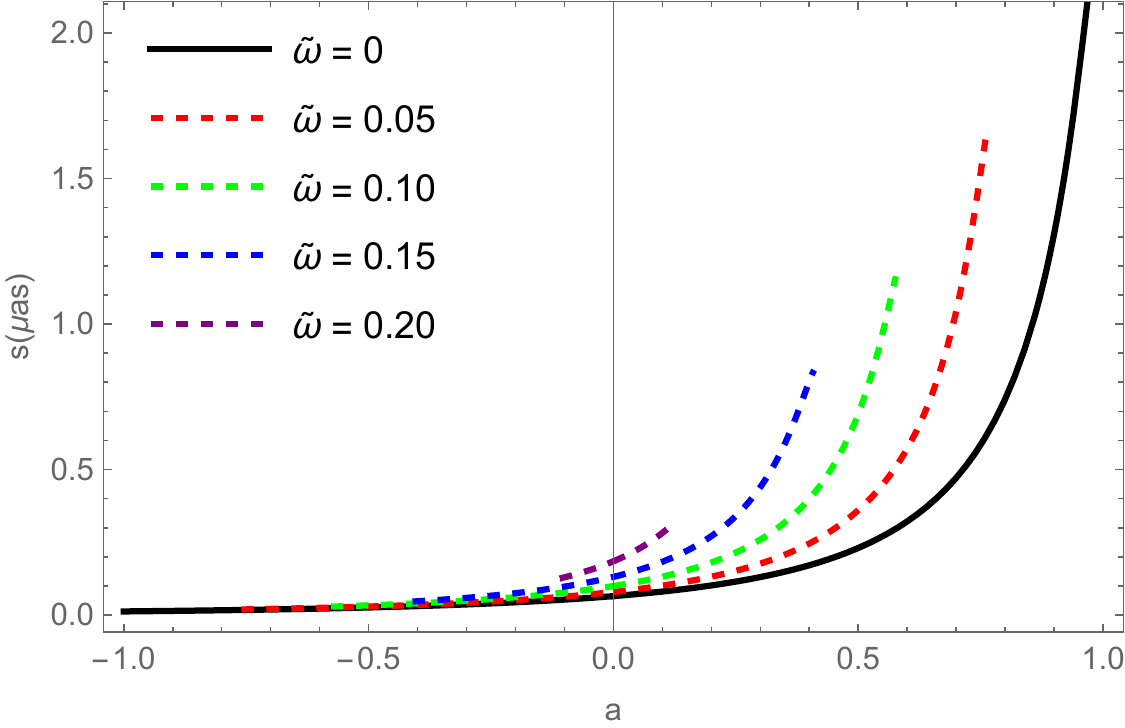}
    \end{subfigure}
    \caption{The behavior of the separations $s$ as a function of $a$ by model M87* (left panel) and Sgr A* (right panel) as the Kerr black hole within QEG.}
    \label{s}
\end{figure*}

So far, we obtain the relationship between the four observables of the relativistic image and the deflection coefficients. Furthermore, we treat M87* and Sgr A* as supermassive Kerr black holes within QEG and evaluate their deviation from the Kerr black hole. In our paper, we take their masses and distances from the Earth as, $M = 6.5\times10^9_\odot$ and $D_{OL} = 16.8$ Mpc for M87* \cite{EventHorizonTelescope:2019dse,EventHorizonTelescope:2019ggy}, as well as $4\times10^6_\odot$ and $D_{OL} =8.0$ kpc for Sgr A* \cite{EventHorizonTelescope:2022wkp,EventHorizonTelescope:2022exc}. However, we directly display the magnification $r_{mag}$ in fig.~\ref{rmag} because $r_{mag}$ does not depend on the mass or distance of the black hole from the observer. The result shows that $r_{mag}$ decreases when we increase $a$. Under the background of M87* (left panel) and Sgr A* (right panel), the image positions $\theta_{\infty}$ are shown in fig.~\ref{theta} as a function of $a$ for different $\tilde{\omega}$. It shows that $\theta_{\infty}$ decreases with the increase of $a$. In fig.~\ref{s}, we plot the separation $s$ under the background of M87* (left panel) and Sgr A* (right panel) between the first image and the packed images. We can see that the increase of $a$ will increase the value of $s$. Compared to the Kerr case, increasing $\tilde{\omega}$ will further reduce the values of both $r_{mag}$ and $\theta_{\infty}$, while $s$ increases with $\tilde{\omega}$ increasing.

\begin{figure*}[t]
\centering
\begin{subfigure}[b]{8.6cm}
        \centering
        \includegraphics[width=8.7cm]{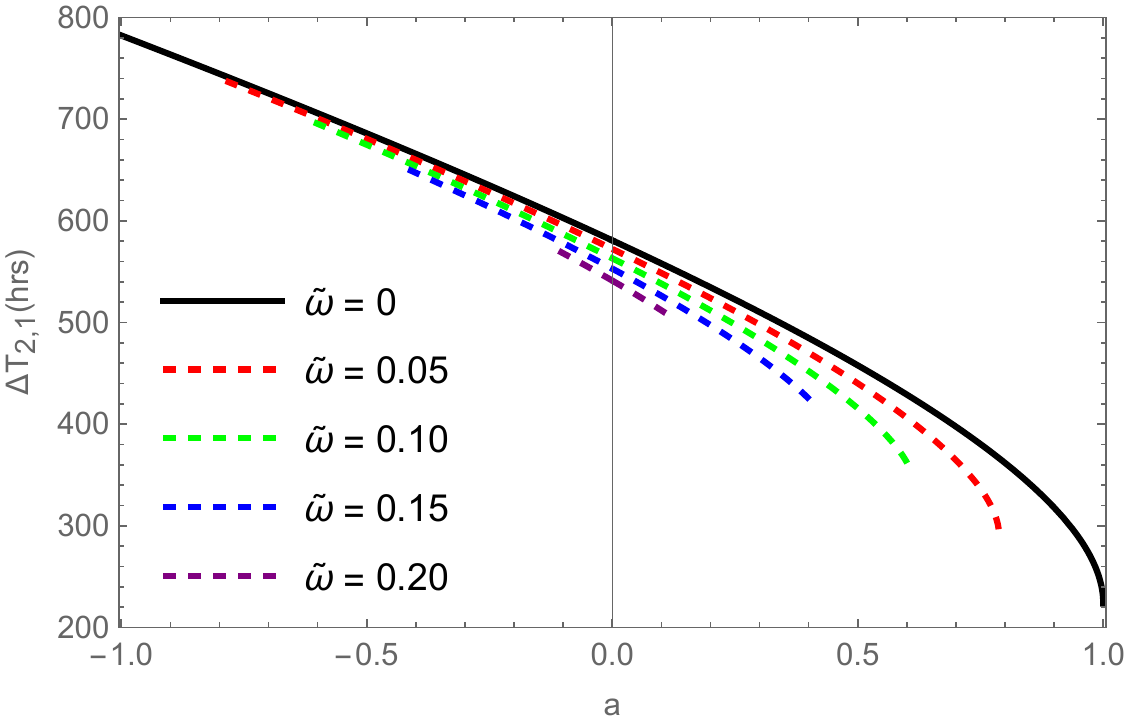}
    \end{subfigure}
    \hspace{0.01\linewidth}
    \begin{subfigure}[b]{8.6cm}
        \centering
        \includegraphics[width=8.5cm]{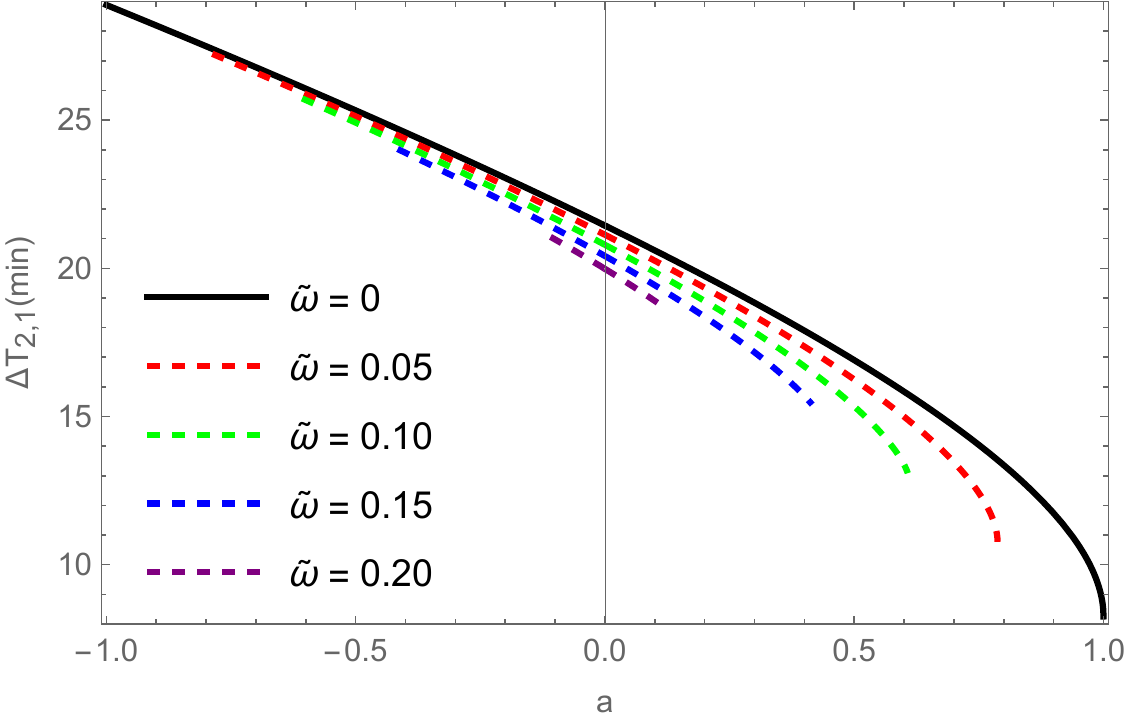}
    \end{subfigure}
    \caption{The behavior of the time delays of first and second images on the same side $\Delta T_{2,1}$ as a function of $a$ by model M87* (left panel) and Sgr A* (right panel) as the Kerr black hole within QEG.}
    \label{ioTD1}
\end{figure*}

By modeling M87* and Sgr A* as the lens, we plot the behaviors of the time delays $\Delta T_{2,1}$ between the first and second images on the same side and between the first and first images on the opposite side $\Delta \widetilde{T}_{1,1}$ in figs.~\ref{ioTD1} and~\ref{ioTD2}. The results show that $\Delta {T}_{2,1}$ decreases for an increase in $a$. For the time delays on the opposite side, $|\Delta \widetilde{T}_{1,1}|$ initially increases with increase in $a$, while $|\Delta \widetilde{T}_{1,1}|$ rapidly changes direction as it approaches an extreme black hole. Although quantum effects limit the value of $a$, they increase $\Delta \widetilde{T}_{1,1}$ compared to the Kerr case when $a$ is fixed. Additionally, for the aforementioned observables, the values modeled with Sgr A* are larger than those with M87*, except for the time delays.

\section{Strong gravitational deflection in plasma}
\label{Strong gravitational deflection in plasma}
In this section, we shall investigate the gravitational lensing of the Kerr black hole within QEG in homogeneous plasma medium, including the effects of plasma on the strong deflection limit coefficients, strong deflection angle, and observables. Let us consider a non-magnetized cold plasma, where the refractive index $n$ of the plasma depends on the spatial location $x^{\alpha}$ and the frequency of the photon ${\omega}(x^\alpha)$ \cite{Bisnovatyi-Kogan:2008qbk}:
\begin{equation}
    n^2=1-\frac{{\omega}_p^2}{[{\omega}(x^\alpha)]^2},\quad{\omega}_p^2=\frac{4\pi e^2N(x^\alpha)}m,
\end{equation}
in which ${\omega}(x^\alpha)$ depends on the spatial coordinates due to gravitational redshift. Here, ${\omega}_p$ is the electron plasma frequency, and $N(x^\alpha)$ is the electron concentration. For homogeneous plasma, ${\omega}_p$ and $N(x^\alpha)$ are constant. It is worth noting that in flat spacetime filled with plasma (dispersive medium) and gravitational fields filled with non-dispersive media, the refractive index $n$ is constant. However, in the gravitational field filled with plasma that we are considering, the refractive index $n$ is not constant. It is affected by gravitational redshift, which specifically manifests as the relationship between the photon frequency and spatial coordinates as follows \cite{Synge:1960ueh}:
\begin{equation}
    p^0\sqrt{-g_{00}}=\hbar {\omega}(x^\alpha).
\end{equation}

The Hamiltonian of photons around the Kerr black hole within QEG filled with homogeneous plasma can be written in the form \cite{Synge:1960ueh}:
\begin{align}
    H(x^i,p_i)&=\frac{1}{2}\bigg[g^{ik}p_ip_k-(n^2-1)\bigg(p_0\sqrt{-g^{00}}\bigg)^2\bigg] \\
    & = \frac12{\left[g^{ik}p_ip_k+\hbar^2 {\omega}_p^2\right]}.
\end{align}

According to the variational principle~(\ref{variational principle}) and the condition $H(x^i,p_i) = 0$, we can derive the set of differential equations \cite{Synge:1960ueh}:
\begin{equation}
    \frac{dx^i}{d\lambda}=\frac{\partial H}{\partial p_i} , \frac{dp_i}{d\lambda}=-\frac{\partial H}{\partial x^i}.
\end{equation}

From these, we can obtain two constants of motion, which represent the energy and angular momentum of the particle:
\begin{equation}
    E=-p_t = \hbar {\omega},\quad L=p_\phi.
\end{equation}

The motion equations of photons in the Kerr spacetime within QEG filled with a uniform plasma are as follows:
\begin{align}
        \frac{dt}{d\lambda}& = \frac{4 E C(r)-2 L D(r) }{D(r)^{2}+4 A(r)C(r)}, \label{dt} \\
        \frac{d\phi}{d\lambda}& = \frac{2 E D(r)+4 L A(r)}{D(r)^{2}+4 A(r)C(r)}, \label{dphi} \\
        \left(\frac{dr}{d\lambda}\right)^{2}& = \frac{1}{B(r)[D(r)^{2}+4 A(r)C(r)]} \notag \\
        & \times \bigg\{4 E^2 C(r)  - \hbar^2 {\omega}^2_p [4A(r)C(r)+D(r)^2] \notag \\
        & - 4 L E D(r)  - 4 L^{2} A(r) \bigg\}. \label{dr}
\end{align}

Since the plasma frequency ${\omega}_{p}$ of homogeneous plasma is constant, we set
\begin{equation}
    \frac{- p_0}{\hbar{\omega}_p}=\frac{{\omega}_0}{{\omega}_p}=\hat E>1,\quad\frac{p_\varphi}{\hbar{\omega}_p}= \hat L > 0.
\end{equation}

Here, ${\omega}_0 = {\omega}(\infty)$ and $\hat{E}$ can indirectly represent the plasma concentration, meaning that a decrease in plasma concentration is equivalent to an increase in $\hat{E}$. It is noteworthy that the spacetime returns to vacuum as $\hat{E} \to \infty$. Correspondingly, an increase in plasma concentration is equivalent to $\hat{E}$ constantly approaching $1$. Additionally, $\hat{E} > 1$ because waves with ${\omega_0} < {\omega}_p$ do not propagate in the plasma.

With the condition $\frac{dr}{d\lambda}\big|_{r=r_0}=0$, we get the angular momentum:
\begin{equation}
    \hat{L}(r_0)=\frac{-\hat{E}D(r_0)+\sqrt{4 A(r_0)PC(r_0) + \hat{E}^2D^2(r_0)}}{2 A(r_0)}
\end{equation}
with
\begin{equation}
    \hat C(r_0)=\hat{E^2}C(r_0) - [A(r_0) C(r_0) + \frac{D(r_0)^2}{4}].
\end{equation}

The impact parameter of photons moving in plasma differs from that in vacuum. Its effective mass is $m_{eff} = \hbar {\omega}_p$ \cite{Kulsrud:1991jt,Bisnovatyi-Kogan:2010flt}, therefore the impact parameter can be expressed as
\begin{equation}
    \bar{u}(r_0) = \frac L{\sqrt{E^2-m_{\mathrm{eff}}^2}}=\frac{\hat{L}}{\sqrt{\hat{E}^2-1}}.
\end{equation}

\begin{figure*}[t]
\centering
\begin{subfigure}[b]{8.6cm}
        \centering
        \includegraphics[width=8.6cm]{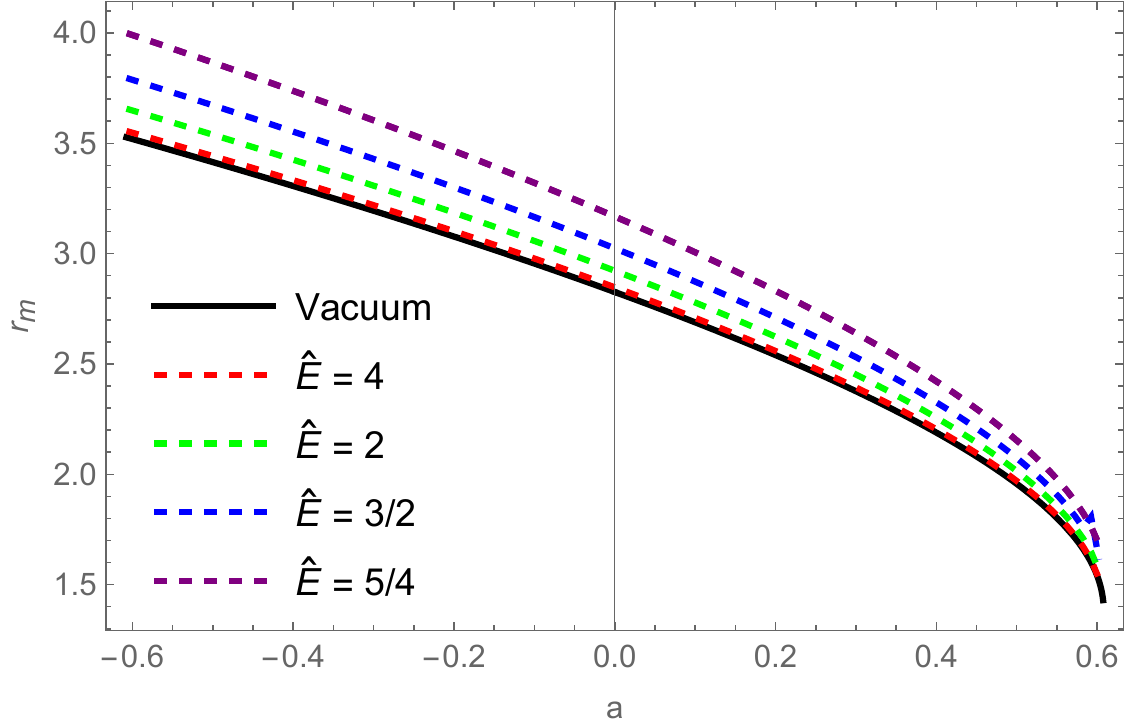}
    \end{subfigure}
    \hspace{0.01\linewidth}
    \begin{subfigure}[b]{8.6cm}
        \centering
        \includegraphics[width=8.6cm]{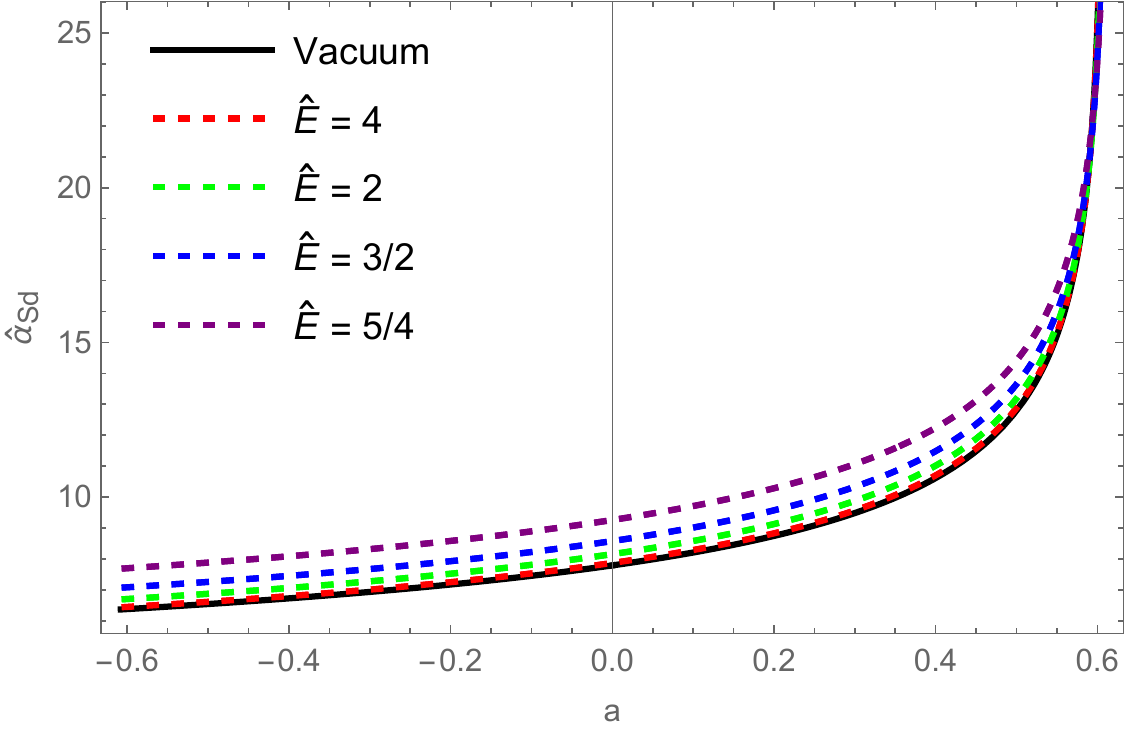}
    \end{subfigure}
    \caption{The left panel: the photon sphere radius $r_m$ as a function of $a$ for the Kerr black hole within QEG in plasma. The right panel: the deflection angle $\hat{\alpha}_{Sd}$ for the Kerr black hole within QEG when $u = u_m + 0.0025$.}
    \label{Rmp}
\end{figure*}

In this spacetime, the equation for the photon sphere reads
\begin{equation}
    A(r) \hat{C}'(r) - A'(r) \hat{C}(r)+ \hat{L} \hat{E} [A'(r) D(r) - A(r) D'(r)] = 0,
\end{equation}
and the radius of the photon sphere for different parameters $\hat{E}$ is shown in the left panel of fig.~\ref{Rmp}. Compared to the vacuum situation, the presence of plasma increases the photon sphere radius $r_m$. Additionally, the strong deflection angle $\hat{\alpha}_{Sd}$ also increases with plasma concentration.

The deflection angle for the photons in the spacetime can be given by eq.~(\ref{angle1}), while $\bar{\phi}(r_{0})$ can be written as
\begin{align}
        \bar{\phi}(r_{0}) =& \int_{0}^{1}\bar{R}(z,r_{0})\bar{f}(z,r_{0})dz, \\
        \bar{R}(z,r_0) =& \frac{2 (1-A(r_0))}{A(r)'} \notag\\
        &\times \frac{\sqrt{B(z)|A(r_0)|}[\hat{E}D(z)+2\hat{L}A(z)]}{\sqrt{\hat{C}(z)}\sqrt{D^2(z)+4A(z)C(z)}}, \\
        \bar{f}(z,r_0) =& \bigg\{sign\boldsymbol{(}A(r_0)\boldsymbol{)}\biggl[A(r_0)-A(z)\frac{\hat{C}(r_0)}{\hat{C}(z)} \notag \\
        & +\frac{\hat{E}\hat{L}}{\hat{C}(z)}(A(z)D(r_0)-A(r_0)D(z))\biggl]\bigg\}^{-\frac{1}{2}},
\end{align}
where $z = \frac{A(r)-A(r_0)}{1-A(r_0)}$.

Similar to the previous section, $\bar{R}(z,r_0)$ is regular for all values of $z$ and $r_0$, and $\bar{f}(z,r_0)$ diverges at $z = 0$. We expand the argument of the square root in $\bar{f}(z,r_0)$ to the second order in $z$ \cite{Bozza:2002af,Liu:2016eju}:
\begin{equation}
    \bar{f}_0(z,r_0)=\frac{1}{\sqrt{\bar{\alpha} z + \bar{\beta}z^2}},
\end{equation}
where
\begin{align}
        &\bar{\alpha}(r_{0}) =~\frac{1-A(r_0)}{A(r_0)' \hat{C}(r_{0}) }\biggl\{A(r_{0})\hat{C}^{\prime}(r_{0})-A^{\prime}(r_{0})\hat{C}(r_{0}) \notag \\
        &+ \hat{E}\hat{L}[A^{\prime}(r_{0})D(r_{0})-A(r_{0})D^{\prime}(r_{0})]\biggr\}, \\
        &\bar{\beta}(r_{0}) = \frac{1-A(r_0)}{2 A(r_0)' \hat{C}^{2}(r_{0})}\boldsymbol{ \Bigg(}2\biggl(\hat{C}(r_{0})- \frac{1-A(r_0)}{A(r_0)'}\hat{C}^{\prime}(r_{0})\biggr) \notag \\
        &\times \biggl\{[A(r_{0})\hat{C}^{\prime}(r_{0})-A^{\prime}(r_{0})\hat{C}(r_{0})]  \notag \\
        &+ \hat{E}\hat{L}[A^{\prime}(r_{0})D(r_{0})-A(r_{0})D^{\prime}(r_{0})]\biggr\}  \notag \\
        &+\frac{1-A(r_0)}{A(r_0)'}\hat{C}(r_0)\bigg\{[A(r_0)\hat{C}''(r_0)-A''(r_0)\hat{C}(r_0)]   \notag\\
        &+ \hat{E}\hat{L}[A''(r_0)D(r_0)-A(r_0)D''(r_0)]\bigg\}\boldsymbol{\Bigg)}. 
\end{align}

Then the strong deflection angle can be obtained by the integral as \cite{Bozza:2002af,Liu:2016eju,Tsukamoto:2016jzh}
\begin{align}
    \hat{\alpha}
    _{Sd}(\theta)=& -\bar{a}_p \log\left(\frac{\theta D_{OL}}{\bar{u}(r_m)}-1\right)+\bar{b}_p \notag \\
    & +\mathcal{O}\boldsymbol{(}(\bar{u}-\bar{u}_m)\log(\bar{u}-\bar{u}_m)\boldsymbol{)},
\end{align}
where
\begin{align}
    \bar{a}_p =& \frac{\bar{R}(0,r_{m})}{2\sqrt{\bar{\alpha}(r_{m})}}, \\
    \bar{b}_p =& -\pi+b_{Rp}+ \bar{a}_p\log[{4\bar{\alpha}(r_{m})\hat{C}(r_{m})}] \notag \\
    & + \bar{a}_p \log \biggl(u_{p}(r_m)\sqrt{\hat{E}^{2}-1}|A(r_{m})| [ \hat{E} D(r_{m}) \notag \\
    & +2 \hat{L}(r_m)A(r_{m})] \biggl), \\
    b_{Rp} = & \int_{0}^{1}[\bar{R}(z,r_{0})\bar{f}(z,r_{0})-\bar{R}(0,r_{0})\bar{f}_{0}(z,r_{0})]dz.
\end{align}

\begin{figure}[t]
\centering
\includegraphics[width=8.6cm]{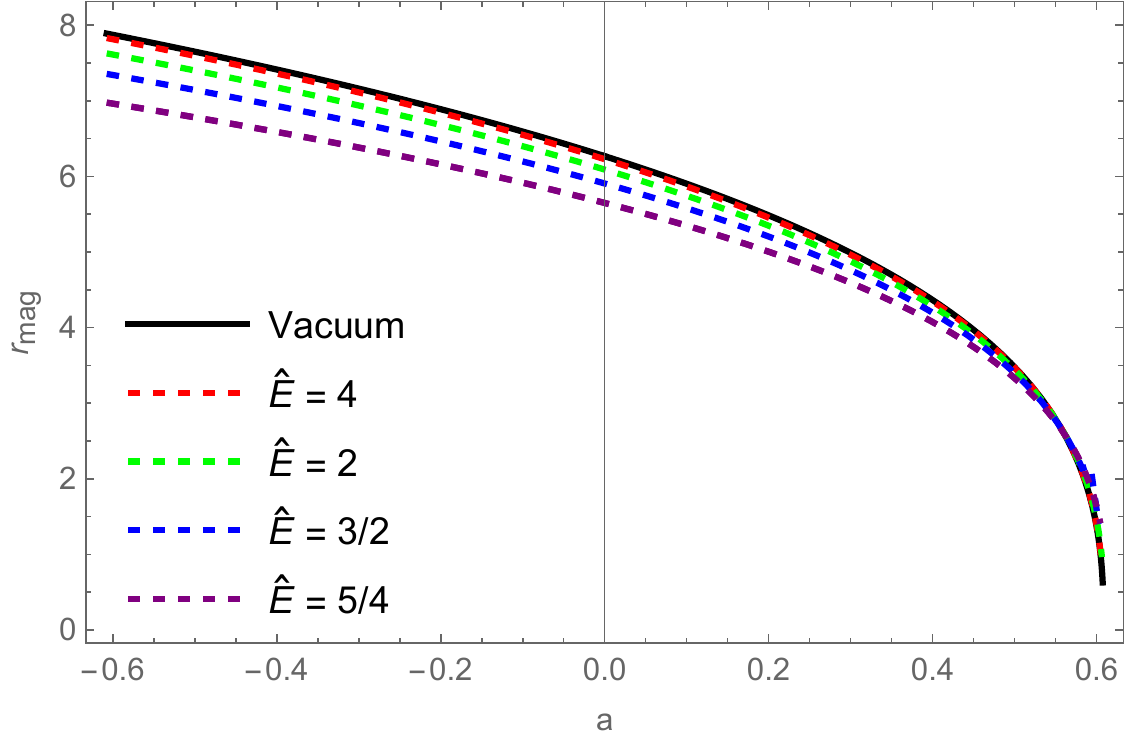}
\caption{The behavior of the magnification $r_{mag}$ as a function of $a$ in plasma.}
\label{rmagp}
\end{figure}

\begin{figure*}[t]
\centering
\begin{subfigure}[b]{8.6cm}
        \centering
        \includegraphics[width=8.6cm]{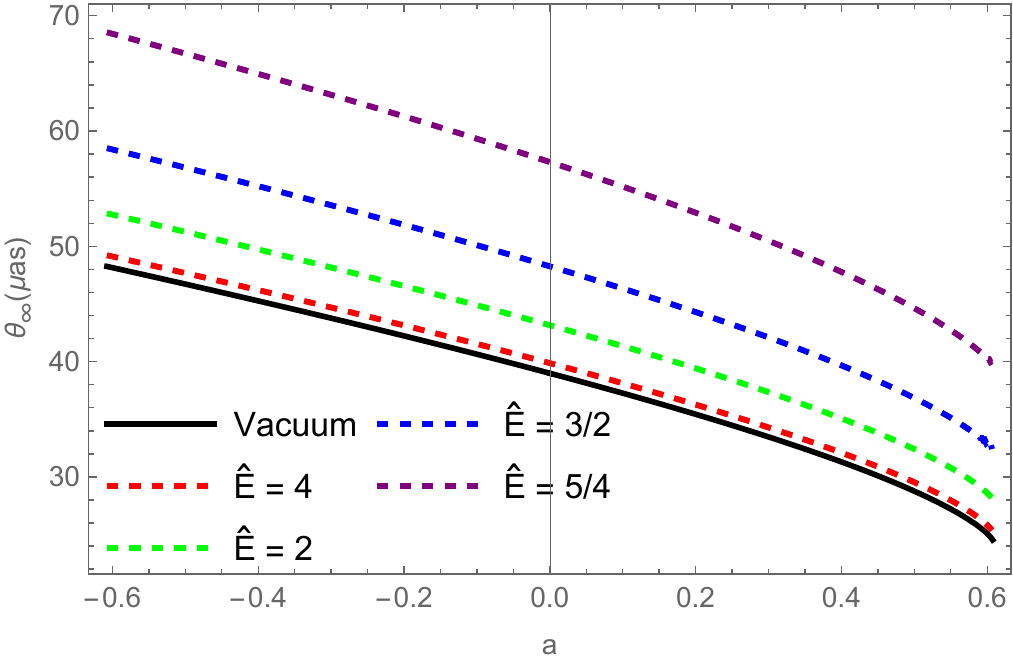}
    \end{subfigure}
    \hspace{0.01\linewidth}
    \begin{subfigure}[b]{8.6cm}
        \centering
        \includegraphics[width=8.6cm]{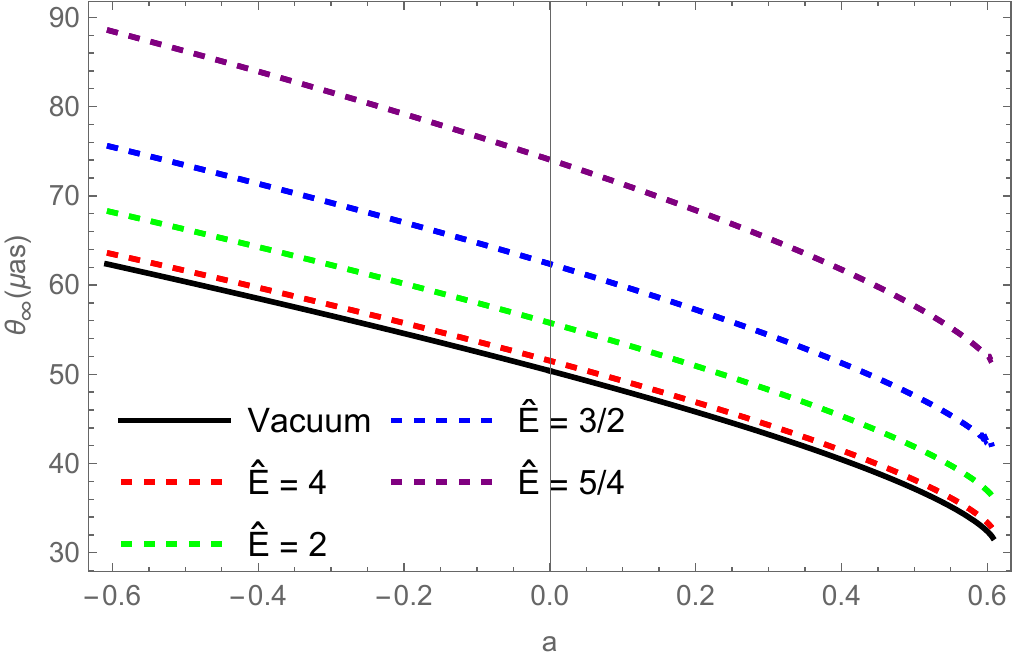}
    \end{subfigure}
    \caption{The image position $\theta_{\infty}$ by modeling M87* (left panel) and Sgr A* (right panel) as the Kerr black hole within QEG in plasma.}
    \label{thetap}
\end{figure*}

\begin{figure*}[t]
\centering
\begin{subfigure}[b]{8.6cm}
        \centering
        \includegraphics[width=8.6cm]{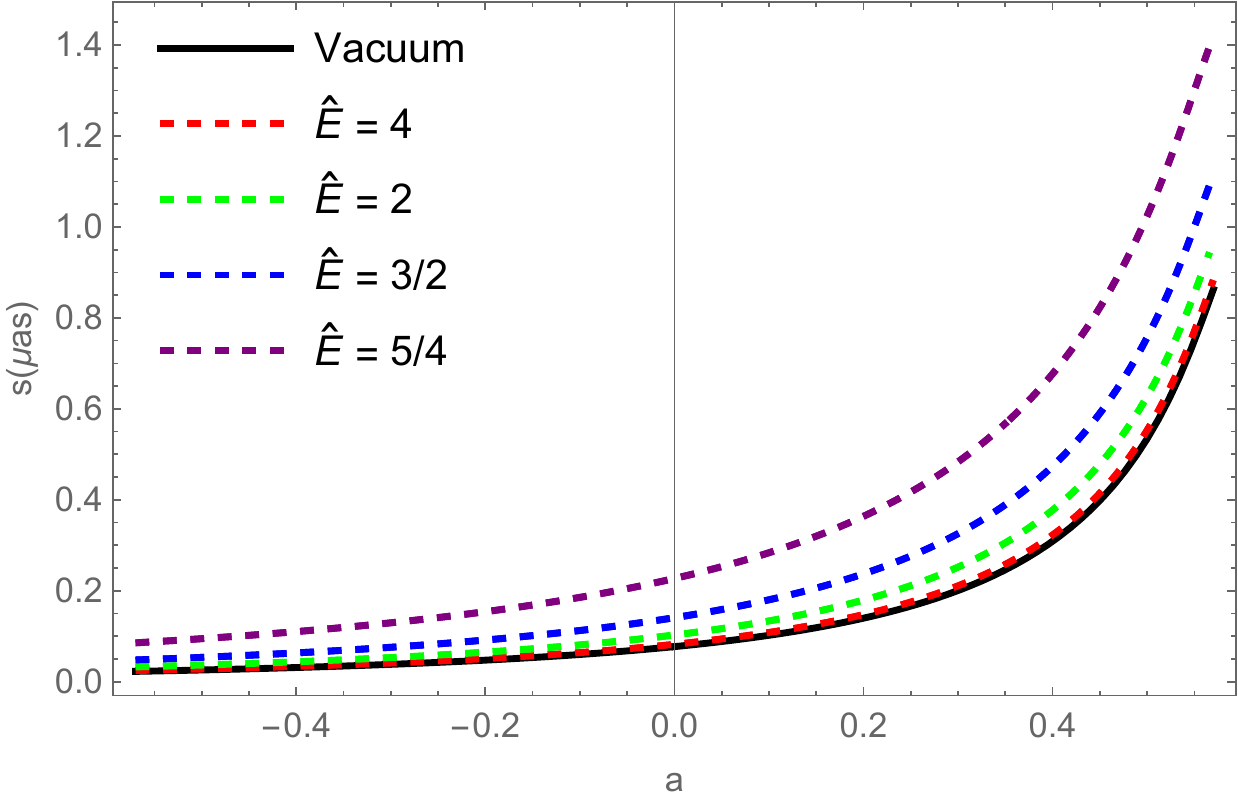}
    \end{subfigure}
    \hspace{0.01\linewidth}
    \begin{subfigure}[b]{8.6cm}
        \centering
        \includegraphics[width=8.6cm]{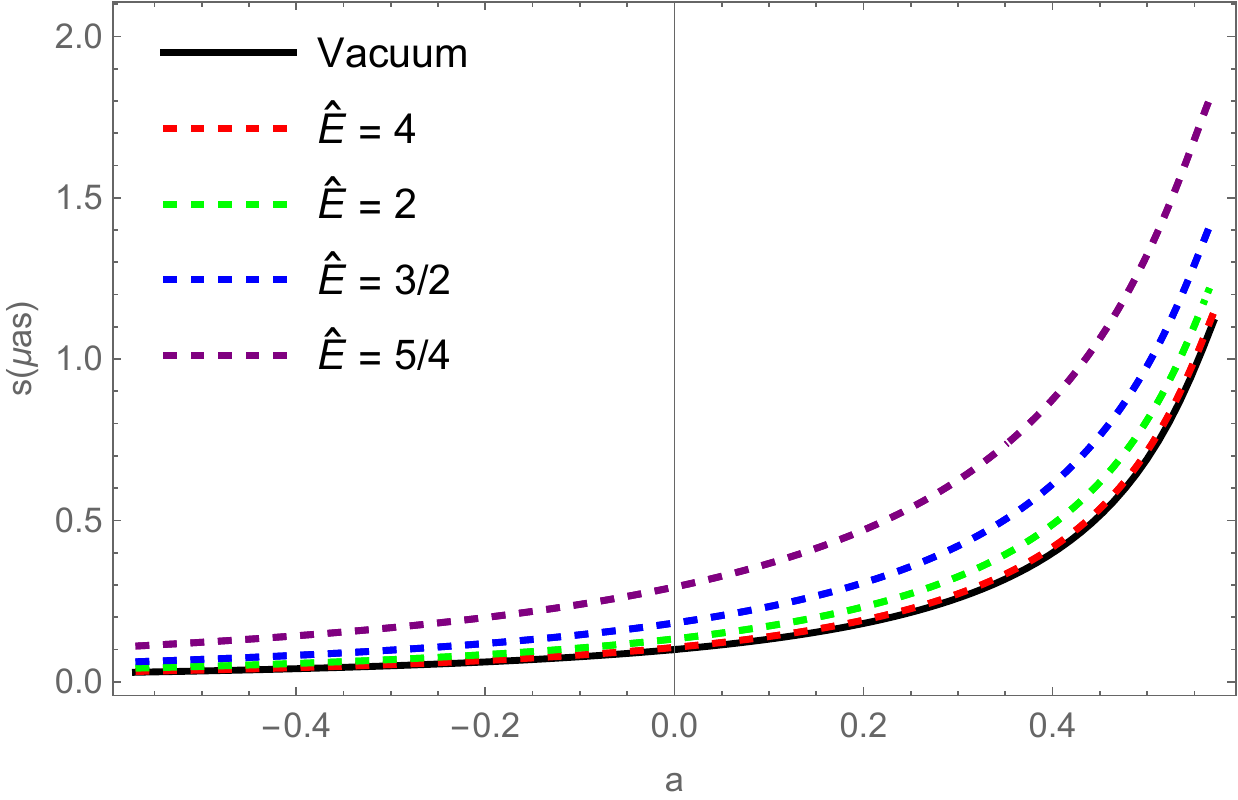}
    \end{subfigure}
    \caption{The separations $s$ by modeling M87* (left panel) and Sgr A* (right panel) as the Kerr black hole within QEG in plasma.}
    \label{sp}
\end{figure*}

Following the pattern shown in section~\ref{Strong gravitational deflection in plasma}, we can obtain the corresponding observables based on the strong deflection coefficient in the plasma situation: the magnification $r_{mag}$ of the $n$th image, the angular position $\theta_{\infty}$ of the asymptotic images, and the separation $s$ between the first image and the others. We also study the behaviors of these observables as a function of $\hat{E}$ when $\tilde{\omega} = 0.1$ and $\gamma = 9/2$. Fig.~\ref{rmagp} shows that $r_{mag}$ decreases with increasing plasma concentration ($\hat{E} \to 1$), while $\theta_{\infty}$ increases with increasing plasma concentration, as shown in fig.~\ref{thetap}. In fig.~\ref{sp}, we plot $s$ as a function of $a$ for different plasma concentrations. The results show that $s$ increases with increasing plasma concentration.

In addition, under the influence of plasma, the expression for the time delays becomes

\begin{align}\label{timedelayp}
        T_{1}-T_{2}=&\widetilde{\mathcal{T}}(r_{0,1})-\widetilde{\mathcal{T}}(r_{0,2})+2\int_{r_{0,1}}^{r_{0,2}}\frac{\widetilde{P}_{2}(r,r_{0,1})}{\sqrt{A_{0,1}}}dr \notag\\
        & +2\int_{r_{0,2}}^\infty\left[\frac{\widetilde{P}_2(r,r_{0,1})}{\sqrt{A_{0,1}}}-\frac{\widetilde{P}_2(r,r_{0,2})}{\sqrt{A_{0,2}}}\right]dr
\end{align}

with
\begin{align}
    \widetilde{\mathcal{T}}(r_0)& =\int_0^1\widetilde{\mathcal{R}}_q(z,r_0) \bar{f}(z,r_0) dz, \label{integralTD} \\
    \widetilde{\mathcal{R}}(z,r_0)& =2\frac{1-A_{0}}{A^{\prime}(r)}\widetilde{P}_{2}(r,r_{0})\left(1-\frac{1}{\sqrt{A_{0}} \bar{f}(z,r_{0})}\right),\\
    \widetilde{P}_{2}(r,r_{0})& =\frac{\sqrt{BA_{0}}[2\hat{E} C - L(r_0) D]}{\sqrt{\hat{C}(r)}\sqrt{4AC+D^{2}}},
\end{align}

Using the same method as in the previous section, we can integrate the result of the integral~(\ref{integralTD}) as follows:
\begin{equation}
    \widetilde{\mathcal{T}}(\bar{u})=-\tilde{a}_p \log\left(\frac{\bar{u}}{\bar{u}_m}-1\right)+\tilde{b}_p+O\left(\bar{u}-\bar{u}_m\right)
\end{equation}
with
\begin{align}
    &\tilde{a}_p=\frac{\widetilde{\mathcal{R}}(0,r_m)}{2\sqrt{\bar{\beta}_m}}, \\
    &\tilde{b}_p = -\pi+\tilde{b}_{D}\left(r_{m}\right)+\tilde{b}_{R}\left(r_{m}\right)+\tilde{a}_p \log\left(\frac{cr_{m}^{2}}{u_{m}}\right) ,\\
    &\tilde{b}_{D}(r_{m}) =2\tilde{a}_p\log\frac{2(1-A_{m})}{A_{m}^{\prime}r_{m}}, \\
    &\tilde{b}_{R}(r_{m}) =\int_{0}^{1}[\widetilde{\mathcal{R}}(z,r_{m})\bar{f}(z,r_{m})-\widetilde{\mathcal{R}}(0,r_{m})\bar{f}_{0}(z,r_{m})]dz.
\end{align}

Now, we get the formulas for calculating the time delays on the same side and the opposite side under plasma, which read as
\begin{align}
    \Delta {\mathcal{T}}_{n,m}^{s} &\approx 2\pi(n-m)\frac{\tilde{a}}{\bar{a}}, \\
    \Delta \widetilde{\mathcal{T}}_{n,m}^o &\approx\frac{\tilde{a}(a)}{\bar{a}(a)}[2\pi n+\beta_{LS}-\bar{b}(a)]+\tilde{b}(a) \notag\\
    &-\frac{\tilde{a}(-a)}{\bar{a}(-a)}[2\pi m-\beta_{LS}-\bar{b}(-a)]-\tilde{b}(-a).
\end{align}

\begin{figure*}[t]
\centering
\begin{subfigure}[b]{8.6cm}
        \centering
        \includegraphics[width=8.68cm]{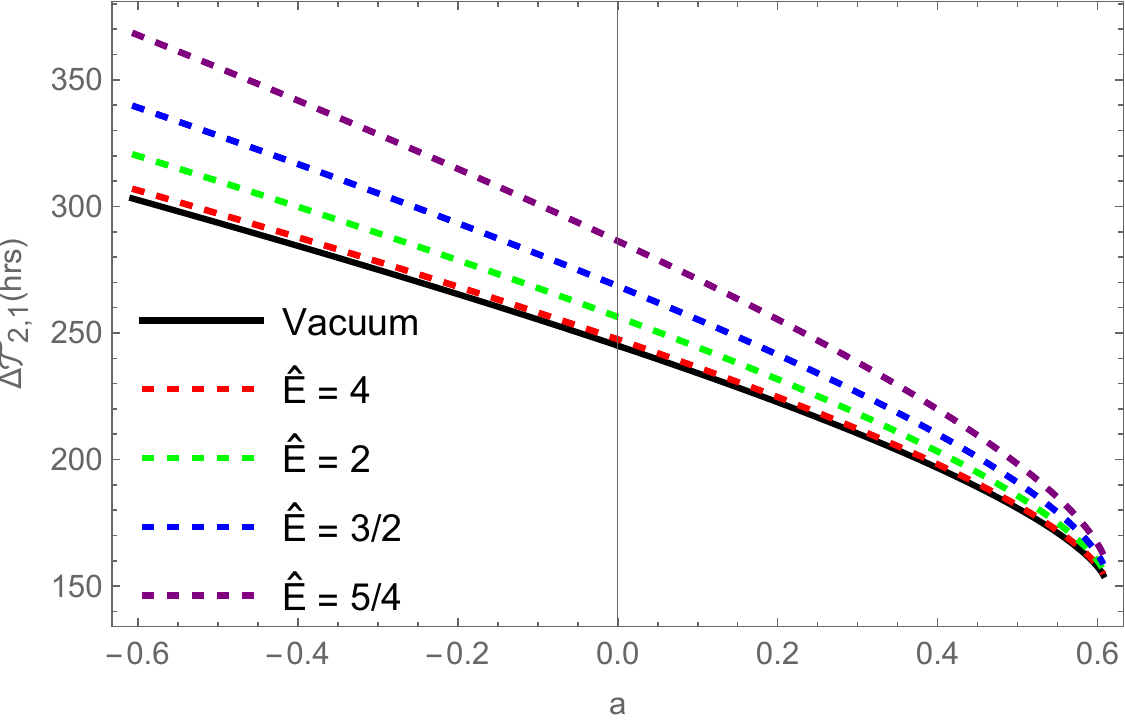}
    \end{subfigure}
    \hspace{0.01\linewidth}
    \begin{subfigure}[b]{8.6cm}
        \centering
        \includegraphics[width=8.6cm]{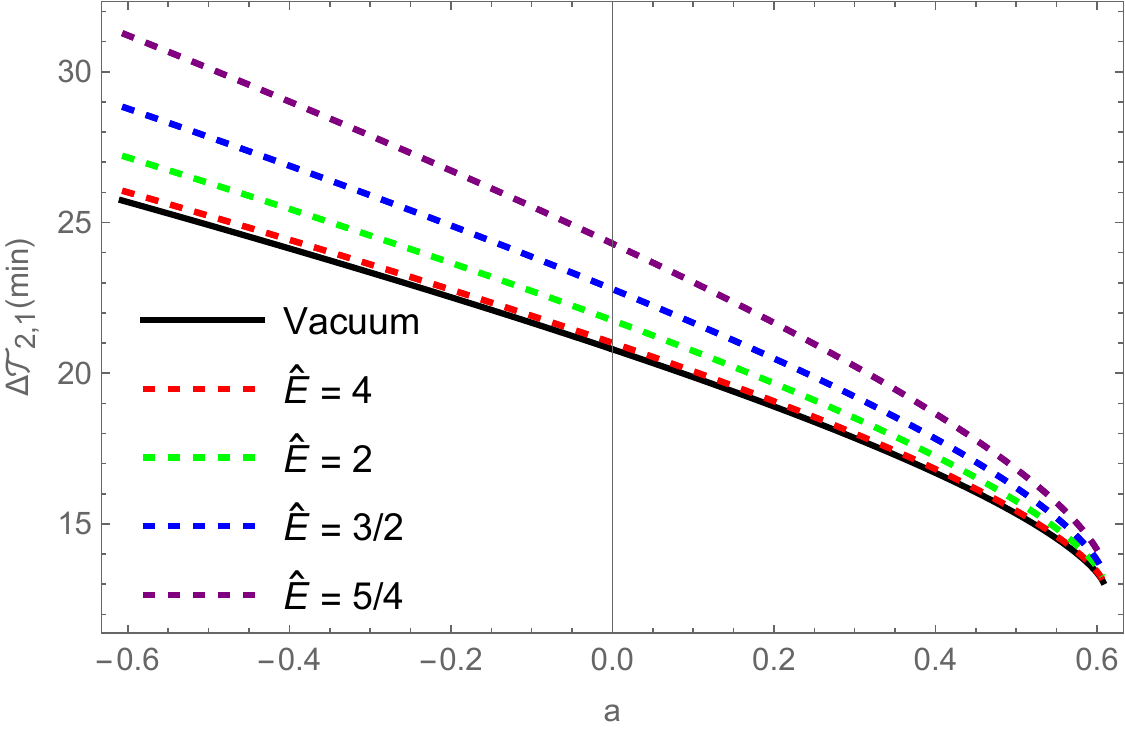}
    \end{subfigure}
    \caption{The time delays on the same side $\Delta {\mathcal{T}}_{2,1}$ by modeling M87* (left panel) and Sgr A* (right panel) as the Kerr black hole within QEG in plasma.}
    \label{TDp}
\end{figure*}

\begin{figure*}[t]
\centering
\begin{subfigure}[b]{8.6cm}
        \centering
        \includegraphics[width=8.7cm]{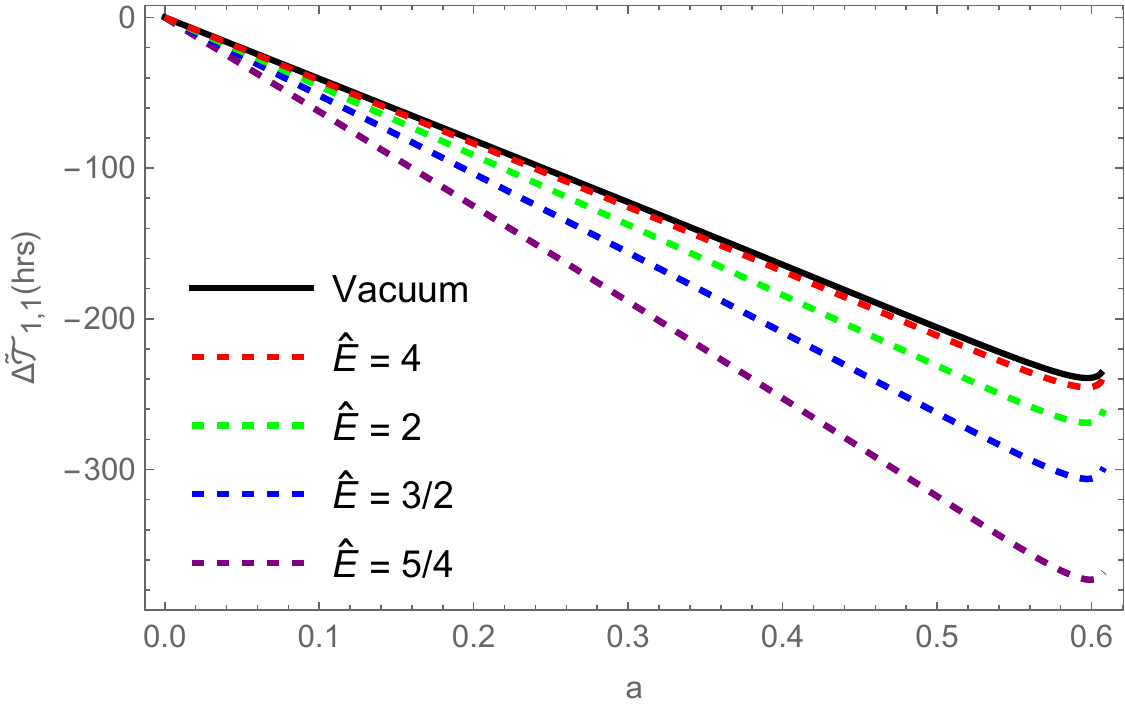}
    \end{subfigure}
    \hspace{0.01\linewidth}
    \begin{subfigure}[b]{8.6cm}
        \centering
        \includegraphics[width=8.6cm]{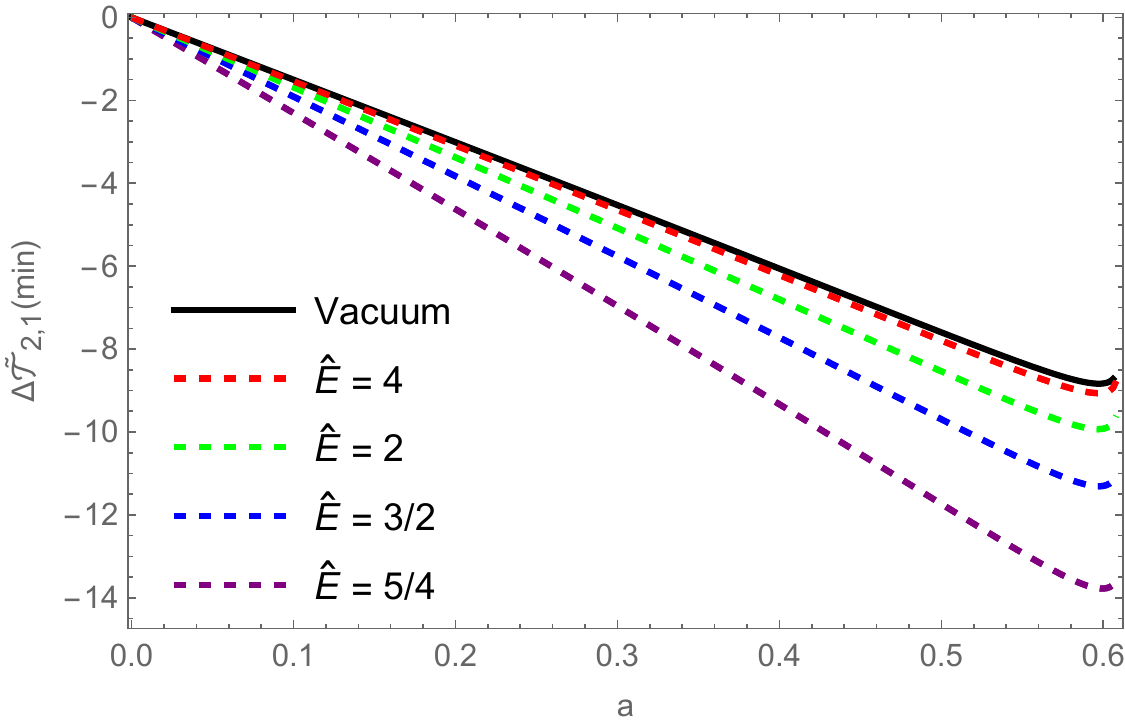}
    \end{subfigure}
    \caption{The time delays on the opposite side $\Delta \widetilde{\mathcal{T}}_{1,1}$ by modeling M87* (left panel) and Sgr A* (right panel) as the Kerr black hole within QEG in plasma.}
    \label{iTDp}
\end{figure*}

Likewise, for the same side, we only consider the dominant term in eq.~(\ref{timedelayp}), while for the opposite side, we consider all terms in eq.~(\ref{timedelayp}). The time delay on the same side $\Delta {\mathcal{T}}_{2,1}$ and the opposite side $\Delta \widetilde{\mathcal{T}}_{1,1}$ for the Kerr black hole within QEG filled with the homogeneous plasma under the background of M87* and Sgr A* is plotted in figs.~\ref{TDp} and~\ref{iTDp}. We can see that $\Delta \widetilde{\mathcal{T}}_{1,1}$ and $|\Delta \widetilde{\mathcal{T}}_{1,1}|$ both increase with an increase in plasma concentration. In addition, the variations of $\Delta \widetilde{\mathcal{T}}_{1,1}$ and $\Delta \widetilde{\mathcal{T}}_{1,1}$ with respect to spin $a$ are the same as in the Kerr case.

\section{Conclusion}
\label{Conclusion}
From the perspective of effective field theory, Bonanno and Reuter regard General Relativity (GR) as the low-energy limit of QEG. Therefore, quantum corrections to the Kerr black hole can be obtained within the framework of QEG. In order to obtain observational signatures of the quantum effects, we have investigated the gravitational lensing effect of the Kerr black hole within QEG in the strong deflection limit. We have considered the propagation of light in the equatorial plane and obtained the radius of the photon sphere, the deflection angle of light, as well as the lensing observables. The affect of the spin parameter $a$ on the black hole is not significantly altered by the quantum effects. Similar to the Kerr-like black hole, with increasing $a$, the photon sphere radius $r_m$ decreases, and the light deflection angle $\alpha_{Sd}$ increases at $u = u_m + 0.0025$. Regarding the influence of the quantum effects, $r_m$ decreases when we increase the size of the quantum effects $\tilde{\omega}$, yet $\alpha_{Sd}$ increases.

To relate to the realistic evidence for the existence of black holes, we treat M87* and Sgr A* as lens models. Moreover, we have calculated the magnification $r_{mag}$, the position of asymptotic relativistic images $\theta_{\infty}$, the separations $s$, and the time delays between different images. Among them, the presence of quantum effects reduces $r_{mag}$, $\theta_{\infty}$, and the time delays on the same side, while increasing $s$ and the time delays on the opposite side. For the observation of the time delays, although the opposite images are more apparent, the time delays on the same side are larger. Therefore, if the observation precision is not sufficient, we can observe the time delays on the same side. Since the time delays on M87* are much greater than on Sgr A*, observing time delays on M87* is a better choice. However, for other observables, they are more significant on Sgr A*. It is worth noting that the behavior of the time delays on the opposite side as a function of $a$ changes dramatically for an extreme black hole.

Furthermore, it is highly likely that plasma exists around black holes, so studying gravitational lensing effects under the influence of plasma is necessary. In the propagation of light, in addition to the deflection caused by gravity, there will be an additional deflection due to the effect of plasma. In this paper, we have computed the photon sphere radius, the strong deflection angle, and observables under spacetime filled with plasma. Especially, we have obtained the time delays under the influence of plasma in the strong deflection limit. The result is that the time delays and other observables increase with the concentration of plasma, except for $r_{mag}$.

To this day, the pursuit of a quantum gravity theory continues. The black hole mentioned in this paper was proposed within the framework of QEG. Our research on the gravitational lensing of this black hole can be used to test the black hole and QEG theory from an astrophysical perspective.


\end{document}